\documentclass[aip,preprint,jmp,author-numerical,floatfix]{revtex4-1}
\usepackage[utf8x]{inputenc}
\usepackage{amsmath,amssymb,graphicx}
\usepackage{url}

\begin{document}
	
\title[Multipole Electrodynamic Ion Trap Geometries for Microparticle Confinement]{Multipole Electrodynamic Ion Trap Geometries for Microparticle Confinement under Standard Ambient Temperature and Pressure Conditions}
	
\author{Bogdan M. Mihalcea} 
\email{bogdan.mihalcea@inflpr.ro}
\affiliation{National Institute for Laser, Plasma and Radiation Physics (INFLPR), \\
Atomi\c stilor Str. Nr. 409, 077125 M\u agurele, Ilfov, Romania}
\author{Liviu C. Giurgiu}
\affiliation{University of Bucharest, Faculty of Physics, Atomistilor Str. Nr. 405, 077125 M\u agurele, Romania}
\author{Cristina Stan}
\affiliation{Department of Physics, {\em Politehnica} University, 313 Splaiul Independen\c tei, RO-060042, Bucharest, Romania}
\author{Gina T. Vi\c san}
\author{Mihai Ganciu}
\affiliation{National Institute for Laser, Plasma and Radiation Physics (INFLPR), \\
Atomi\c stilor Str. Nr. 409, 077125 M\u agurele, Ilfov, Romania}
\author{Vladimir E. Filinov}
\author{Dmitry Lapitsky}
\email{dmitrucho@yandex.ru}
\author{Lidiya Deputatova}
\author{Roman Syrovatka}
\affiliation{Joint Institute for High Temperatures, Russian Academy of Sciences, Izhorskaya Str. 13, Bd.  2, 125412 Moscow, Russia}

\begin{abstract}
Trapping of microparticles and aerosols is of great interest for physics and chemistry. We report microparticle trapping in case of multipole linear Paul trap geometries, operating under Standard Ambient Temperature and Pressure (SATP) conditions. An 8-electrode and a 12-electrode linear trap geometries have been designed and tested with an aim to achieve trapping for larger number of particles and to study microparticle dynamical stability in electrodynamic fields. We report emergence of planar and volume ordered structures of microparticles, depending on the a.c. trapping frequency and particle specific charge ratio. The electric potential within the trap is mapped using the electrolytic tank method. Particle dynamics is simulated using a stochastic Langevin equation. We emphasize extended regions of stable trapping with respect to quadrupole traps, as well as good agreement between experiment and numerical simulations.   
\end{abstract}



\maketitle
	
\section{Introduction}
\label{intro}

Coulomb systems represent many-body systems made of identical particles, that interact by means of electrostatic forces. When the potential energy associated to the Coulomb interaction is larger than the kinetic energy of the thermal motion, the system is strongly coupled as it presents a strong spatial correlation between the electrically charged particles, similar to liquid or crystalline structures. Strongly coupled Coulomb systems encompass various many-body systems and physical conditions, such as dusty (complex) plasmas or non-neutral and ultracold plasmas \cite{Davidson2001, Werth2005a, Tsyto2008, Mendonca2013, Ott2014, Knoop2015}. Complex plasmas represent a unique type of low-temperature plasmas, characterized by the presence of highly charged nano or microparticles, by chemical reactions, and by the interaction of plasmas with solid surfaces \cite{Bonitz2010a}. They are encountered in astrophysics as interstellar dust clouds, in comet tails or as spokes in the ring systems of giant gas planets \cite{Fortov2010, Shukla2002}. Complex plasmas are also present in the mesosphere and troposphere of the Earth, near artificial satellites and space stations, or in laboratory experiments \cite{Morfill2009, Fortov2010, Bonitz2014}. Dust particle interaction occurs via shielded Coulomb forces \cite{Shukla2002b}, the so-called Yukawa interaction \cite{Fortov2010}. Yukawa balls were reported in case of harmonically confined dusty plasmas \cite{Bonitz2010a, Trevi2006, Bonitz2010b}. The dynamical time scales associated with trapped microparticles lie in the tens of milliseconds range, while microparticles can be individually observed using optical methods \cite{Davis2002, Libbrecht2015}. As the background gas is dilute, particle dynamics exhibits strong coupling regimes characterized by collective motion \cite{Davidson2001, Werth2005a, Tsyto2008, Fortov2007}. Dust particles may give birth to larger particles which might evolve into grain plasmas \cite{Tsyto2008}. The mechanism of electrostatic coupling between the grains can vary widely from the weak coupling (gaseous) regime to the pseudo-crystalline one \cite{Werth2005a, Dubin1999, Lisin2013}. Complex plasmas can be described as non-Hamiltonian systems of few or even many-body particles. They are investigated in connection with issues regarding fundamental physics such as phase transitions, self-organization, study of classical and quantum chaos, pattern formation and scaling. 

The paper investigates strongly coupled Coulomb systems of finite dimensions, such as electrically charged microparticles confined in electrodynamic traps. Particular examples of such systems would be electrons and excitons in quantum dots \cite{Bonitz2014, Chen2007} or laser cooled ions confined in Paul or Penning type traps \cite{Davis2002, Bolli2003, Major2005, Haroche2013, Quint2014, Knoop2014}. We can also mention ultracold fermionic or bosonic atoms confined in traps or in the periodic potential of an optical lattice \cite{Bonitz2010b}. Experimental investigations of charged particles in external potentials have gained from the invention of ion traps \cite{Paul1958, Paul1990}, as they have greatly influenced the future of modern physics and state of the art technology \cite{Knoop2015, Major2005, Ghosh1995, Werth2009}. Using a classical (hyperbolic) Paul trap geometry in vacuum, ordered structures of iron and aluminium microparticles with micron range diameter have been observed since the 1960s \cite{Wuerker1959}. The particles repeatedly crystallized into a regular array and then melted owing to the dynamical equilibrium between the trapping potential and inter-particle Coulomb repulsion. In 1991 an experiment demonstrated the storage of macroscopic dust particles (anthracene) in a Paul trap \cite{Winter1991}, as friction in air was proven to be an efficient mechanism to {\em cool} the microparticles. The mechanism is similar with cooling of ions in ultrahigh vacuum conditions owing to collisions with the buffer gas molecules \cite{Major2005}. The dynamics of a charged microparticle in a Paul trap near Standard (Ambient) Temperature and Pressure conditions was studied by Izmailov \cite{Izmailov1995}, using a Mathieu differential equation with damping term and stochastic source, under conditions of combined periodic parametric and random external excitation. Particle dynamics in nonlinear traps has also been investigated. It was found that ion motion is well described by the Duffing oscillator model with an additional nonlinear damping term \cite{Mihalcea2010, Akerman2010}. Regions of stable (chaos) and unstable dynamics were illustrated, as well as the occurrence of strange attractors and fractal properties for the associated dynamics. 

The Paul trap proved to be a very powerful tool to look into the physics of few-body phase transition since its very beginning \cite{Blumel1990, Walther1995, Schlipf1996}. High-precision spectroscopy and mass spectrometry measurements with unprecedented accuracy \cite{Knoop2015, Quint2014, Margolis2004, Rosen2007, Bushev2013}, quantum physics tests \cite{Haroche2013, Werth2009}, precise control of quantum states \cite{Leibf2003, Home2011}, study of non-neutral plasmas \cite{Dubin1999, Bolli2003, Major2005, Knoop2014}, optical frequency standards \cite{Peik2006, March2010, Ludlow2015}, quantum metrology and quantum information processing (QIP) experiments \cite{Quint2014, Knoop2014, Chiav2005, Leibf2007, Zutic2004, Kim2005} and use of optical transitions in highly charged ions for detection of variations in the fine structure constant \cite{Knoop2015, Quint2014, King2012}, became all possible by using laser cooled atomic and molecular ions confined in electrodynamic and Penning traps \cite{Knoop2014}. 

The paper is organized as follows: Section \ref{intro} introduces Coulomb systems (especially strongly coupled examples of such systems), as they encompass various many-body systems and physical conditions. Because microparticles can be found as atmospheric aerosols or astrophysical dusty plasmas, investigation of such mesoscopic systems requires high precision mass measurements for micro and nanoparticles. Section \ref{micro} reviews different investigation methods for micro and nanoparticles, as their presence in the atmosphere determines the quality of life. Section \ref{microdyn} reviews the equations that govern aerosol (microparticle) dynamics and the forces implied. The equation describing trapped particle dynamics in air is introduced, using dimensionless variables. The trap geometries under test are presented in Section \ref{exp}, together with experimental parameters and the electronic setup. The transversal section of the electric field is mapped for the 12-electrode trap. Section \ref{results} illustrates the occurrence of ordered and stable structures of microparticles within the trap. A physical modelling is performed using a stochastic Langevin equation to account for microparticle dynamics in Section \ref{model}. The equation that characterizes the trap potential is inferred, for experimental trap parameters of interest. Regions of stable trapping are illustrated and microparticle trajectories for the two trap geometries are investigated, by means of numerical simulations. Section \ref{conclusions} emphasizes compatibility with previous experiments (results) as well as the good agreement between numerical theory and experiment.

\section{Microparticles. Investigation Methods}\label{micro}

Microparticles can be found as atmospheric aerosols \cite{Pandis1995, Seinfeld2006}, astrophysical dusty plasmas \cite{Bonitz2014, Fortov2007, Draine2003, Fortov2011} or plasma membranes in case of biological cells and bacteria. Recent investigations have demonstrated a strong correlation between the presence of aerosols in the atmosphere and their effect on climate parameters such as local temperature, air quality and rainfalls, things directly related to the phenomenon of global warming and the quality of life \cite{Trevi2006, Seinfeld2006, Kulkarni2011}. Aerosol particles with diameter less than 10 $\mu$m enter the pulmonary bronchi, while those whose diameter is lower than 2.5 $\mu$m reach the pulmonary alveoli, a region where gas exchange takes place. The presence of certain aerosols (especially anthropogenic ones, such as smoke, ashes or dust) is associated with high levels of industrial pollution and it is responsible for respiratory and cardiovascular diseases, as well as for the ever increasing incidence of human allergies in town areas. Investigation of atmospheric aerosols, viruses, bacteria, and chemical agents responsible for environment pollution, requires high precision mass measurements for micro and nanoparticles (with dimensions ranging between 10 - 10,000 nm), in order to characterize and explain the underlying chemistry and physics of such complex systems \cite{Seinfeld2006, Kulkarni2011, Wester2009}. Moreover, study of such mesoscopic systems \cite{Walther1995} is of large interest, as mesoscopic physics is linked to the fields of nanofabrication and nanotechnology. 

Since its early days the Paul trap \cite{Knoop2015, Paul1990} has proven to be a versatile device that uses path stability as a means of separating ions according to their specific charge ratio \cite{Kaiser1991, Schwartz1991, March2005}. Ion dynamics in Paul traps is characterized by linear, uncoupled equations of motion (Mathieu equations), that can be solved analytically \cite{Major2005, Knoop2014, Ghosh1995}. The Quadrupole Ion Trap Mass Spectrometry (QIT-MS) is a promising technique to perform mass analysis of micron-sized particles such as biological cells, aerosols and synthetic polymers \cite{Davis2002}. The trap can be operated in the mass-selective axial instability mode by scanning the frequency of the applied a.c. field \cite{Nie2008}. When using the QIT as a electrodynamic balance to perform microparticle diagnosis, the frequency of the a.c. field applied to the trap electrodes is typically in the range of 1 kHz or less. Because of this low frequency, in order to achieve high mass measurement accuracy (better than $1$ ppm) only one particle is analyzed at a time, over a time period ranging from seconds up to minutes.

Fluorophore labeled polystyrene microparticles (with dimensions ranging between 2 to 7 microns) have been trapped in vacuum, by means of three dimensional electrodynamic quadrupolar fields and probed using laser fluorescence spectroscopy and QIT-MS. Investigation of single particle emission spectra and of associated optical resonance signatures by using the Mie theory, supplies information regarding the  particle  size, shape and its refractive  index \cite{Trevi2006, Trevi07}. The absolute  mass  and electric charge  of  single  microspheres  can be determined  by  measuring their secular oscillation frequencies. Moreover, a microsphere can act as a three-dimensional optical cavity able to support optical resonances, also known as morphology-dependent resonances (MDR). Using individual micron-sized particles doped with fluorescence dyes and confined within a quadrupole trap, MDR emission has been observed  by means of confocal and two-photon fluorescence microscopes \cite{Smith2008}. Optical MDR resonances induced in the fluorescence spectrum were investigated in order to study laser-induced coalescence of two conjoined, polystyrene spheres levitated in a QIT \cite{Trevi09}.

Thus, QIT-MS is an excellent tool that can be used to identify organic molecules in complex samples. A specific charge $m/z$ can be isolated in the ion trap by ejecting all other $m/z$ particles (ions) by applying various resonant frequencies \cite{Nie2008, Stoican2008}. Moreover, an ion trap can be coupled to an Aerosol Mass Spectrometer \cite{Wang2006, Kurten2007} to investigate atmospheric aerosol (nano)particles.

To minimize both micromotion and second-order Doppler shift owing to space charge repulsion of ions from the trap node line, multipole electrodynamic (Paul) ion traps have been investigated, where ions are weakly bound with confining fields that are effectively zero through the trap interior and grow rapidly near the trap electrode walls \cite{Wester2009, Gerlich2008a, Gerlich2008b}. Multipole traps are used as tools in analytical chemistry \cite{Gerlich1992, Trippel2006}. Octupole RF traps are used as buffer zones or guides in order to transport ions from one region to another in similar applications. Because ions in a multipole Paul trap spend relatively little time in the region of high RF electric fields, the RF heating phenomenon is greatly reduced and low-temperature ion-molecule collisions can be studied in a well-controllable environment \cite{Wester2009, Otto2009}. This immunity to changes in the ion number comes together with a reduced sensitivity to changes in the ambient temperature. Space-charge effects are not negligible in this situation. It is supposed that optical frequency standards based on multipole traps should be able to achieve sensibly higher accuracy, required by applications such as satellite-based navigation (GPS), quantum metrology or precision measurements on variations of the fundamental constants in physics \cite{Knoop2015, Knoop2014, Burt2006, March2010, Ludlow2015}.

\section{Microparticle Dynamics. Strongly Coupled Coulomb Systems}\label{microdyn}
Dust particles experience a certain number of forces. The prevailing forces acting upon the micrometer sized particles of interest are the a.c. electric trapping field (whose outcome is the ponderomotive force caused by particle motion within a strongly nonlinear electric field) and gravity. The gravitational force can be expressed as \cite{Bonitz2010a}
\begin{equation}
\mathbf{F}_g = m \mathbf{g} = \frac{4}{3} \pi a^3 \rho_d \mathbf{g} \ ,
\end{equation}    
where $\mathbf{g}$ stands for the gravitational acceleration, $m$ represents the dust mass, $a$ is the dust particle radius and $\rho_d$ corresponds to the dust particle density. The electric field yields a force 
\begin{equation}
\mathbf{F}_{el} = -Q\mathbf{e} = -Ze\mathbf{E} \ ,
\end{equation} 
where $Q = Ze$ is the dust particle electric charge and $e = 1.6\cdot10^{-19}$ C is the electron electric charge. When a temperature gradient is created within the neutral gas background, a force called thermophoretic drives the particles towards regions of lower gas temperature. Another force acting upon the particle is the ion drag force, owing to a directed ion flow \cite{Tsyto2008, Fortov2007, Vasilyak2013}. To resume, the forces that act on an aerosol particle include the radiation pressure force, the thermophoretic force, the photophoretic force, electric forces, and possibly magnetic forces, in addition to aerodynamic drag and the force of gravity \cite{Tsyto2008}. 

A one-component plasma (OCP) consists of a single species of charge submerged in the neutralizing background field \cite{Fortov2007, Dubin1999}. Single component non-neutral plasmas confined in Penning or Paul (RF) traps exhibit oscillations and instabilities owing to the occurrence of collective effects \cite{Werth2005a}. The dimensionless coupling parameter that describes the correlation between individual particles in such plasma can be expressed as
\begin{equation}
\Gamma = \frac{1}{4\pi \epsilon_0 } \frac{q^2}{a_{WS} k_B T} \ ,
\end{equation}      
where $\epsilon_0$ is the electric permittivity of free space, $q$ stands for the ion charge, $a_{WS}$ is the Wigner-Seitz radius, $k_B$ denotes the Boltzmann constant and $T$ is the particle temperature. The Wigner-Seitz radius results from the relation
\begin{equation}
\frac{4}{3}\pi {a}^3_{WS} = \frac{1}{n} ,
\end{equation}  
where $n$ is the ion (particle) density. Although the Wigner-Seitz radius measures the average distance between individual particles, it does not coincide with the average inter-particle distance. The $\Gamma$ parameter represents the ratio between the potential energy of the nearest neighbour ions (particles) and the ion thermal energy. It describes the thermodynamical state of an OCP. Low density OCPs can only exist at low temperatures. Any plasma characterized by a coupling factor $\Gamma > 1$ is called strongly coupled. For $\Gamma < 174$  the system exhibits a liquid-like structure, while larger values might indicate a liquid-solid phase transition into a pseudo-crystalline state (lattice) \cite{Werth2005a, Winter1991}. OCPs are supposed to exist in dense astrophysical objects. An example of a high temperature system would be a quark-gluon plasma (QGP), used to characterize the early Universe and  ultracompact matter found in neutron or quark stars \cite{Bonitz2010b, Fortov2011}.  

Colloidal suspensions of macroscopic particles and complex plasmas represent other examples of strongly coupled Coulomb systems. Trapped micrometer sized particles interact strongly over long distances, as they carry large electrical charges. Friction in air combined with large microparticle mass results in an efficient particle {\em cooling}, which leads to interesting strong coupling features. The signature of such phenomenon lies in the appearance of ordered structures, liquid or solid like, as phase transitions occur in case of such systems \cite{Davidson2001, Fortov2007, Dubin1999, Major2005, Knoop2014}. The presence of confining fields maintains the particles localized together in an OCP. Trapped and laser cooled ions offer a good, low temperature realization of a strongly coupled ultracold laboratory plasma. 

The equation of motion for a particle trapped in air, can be expressed in a convenient form by introducing dimensionless variables defined as $Z = z/z_0$ and $\tau = \Omega t/2$, where $z_0$ represents the trap radial dimension (a geometrical constant) and $\Omega $ stands for the frequency of the a.c. trapping voltage. The nondimensional equation of motion then becomes \cite{Kulkarni2011}
\begin{equation}
\frac{d^2 Z}{dt^2} + \gamma \frac{dZ}{dt} +2\beta Z \cos(2\tau) = \sigma \ ,
\end{equation}
where $\gamma$ stands for the drag parameter, $\beta$ is the a.c. field strength parameter and $\sigma$ represents a d.c. offset parameter, defined as   
\begin{equation}
\gamma =\frac{6 \pi \mu d_p \kappa}{m\Omega} \ , \beta = \frac{4g}{\Omega^2}\left(\frac{V_{ac}}{V_{dc}} \right) \ , \sigma = -\frac{4g}{\Omega^2 z_0}\left(\frac{V_{dc}}{V^*_{dc}}\right) ,
\end{equation}\label{extforce}
where $ V^*_{dc} $ satisfies
\begin{equation}
F_z - mg = qC_0\frac{V^*_{dc}}{z_0} .
\end{equation}
$C_0$ and $C_1$ are two geometrical constants ($ C_0<1 $) and $ b = z_0C_0/C_1 $. For a negative charged particle, the right hand term of Eq. \ref{extforce} is a positive quantity. When the d.c. potential is adjusted to compensate the external vertical forces, $V_{dc} = V^*_{dc}$, $\sigma =0$ and the particle experiences stable confinement. Other relevant quantities are the gas viscosity coefficient $\mu$, the particle diameter $d_p$, the particle mass $m$, while $b$ represents the geometrical constant of the electrodynamic balance.

\section{Experimental setup}\label{exp}

\subsection{Microparticle physics using quadrupole and multipole traps}

A linear Paul trap uses a combination of time varying and static electric potentials to create a trapping configuration that confines charged particles such as ions, electrons, positrons, micro or nanoparticles \cite{Knoop2015, Orszag2016}. A radiofrequency (RF) or a.c. voltage is used in order to generate an oscillating quadrupole potential in the $y-z$ plane, which achieves radial confinement of the trapped particles. Axial confinement of positively charged ions (particles) is achieved by means of a static potential applied between two endcap electrodes situated at the trap ends, along the trap axis ($x$ plane). The RF or a.c. field induces an effective potential which harmonically confines ions in a region where the field exhibits a minimum, under conditions of dynamic stability \cite{Knoop2015, Ghosh1995, Major2005, March2005,Vinitsky2015, Libbrecht2015}. As it is almost impossible to achieve quadratic a.c. and d.c. endcap potentials for the whole trap volume, it can be assumed that the potential in the vicinity of the trap axis region can be regarded as harmonic, which is a sufficiently accurate approximation.

Stable confinement of a single ion in the radio-frequency (RF) field of a Paul trap is well known, as the Mathieu equations of motion can be analytically solved \cite{Knoop2015, Ghosh1995, Major2005, March2005}. Particle dynamics in a multipole trap is quite complex, as it is described by non-linear, coupled, non-autonomous equations of motion. Solutions of such system can only be found by performing numerical integration \cite{Fortov2010, Major2005, March2005, Riehle2004}. The assumption of an effective trapping potential still holds, and ion dynamics can be separated into a slow drift (the {\em secular motion}) and the rapid oscillating micromotion \cite{Otto2009, Champenois2009}. The effective potential in case of an ideal cylindrical multipole trap can be expressed as

\begin{equation}
V^{*}(r) = \frac{1}{4} \frac{n^2(qV_{ac})^2}{m \Omega^2 {r_0}^2}{\left(\frac{r}{r_0}\right)}^{(2n - 2)} \ ,
\end{equation}         
where $V_{ac}$ is the amplitude of the radiofrequency field and $n$ is the number of poles (for example $n=2$ for a quadrupole trap, $n = 4$ for an octupole trap and $n = 6$ for a 12-pole trap). The larger the value of $n$, the flatter the potential created within the trap centre and the steeper the potential close to the trap electrodes. 

Further on we will review some of the most important milestones that describe experiments with microparticles in quadrupole traps and experiments with multipole traps, in an attempt to describe current status in the field. An electrodynamic trap, used for investigating charging processes of a single grain under controlled laboratory conditions was proposed in \cite{Beranek2010}. A linear cylindrical quadrupole trap was used, where every electrode was split in half in order to achieve a harmonic trapping potential and thus perform precise measurements on the specific charge-to-mass ratio. The secular frequency of the grain was measured in order to determine the charge to mass ratio, a method previously applied in case of microparticles \cite{Stoican2008}. In the paper of Vasilyak \cite{Vasilyak2013} and co-workers, mathematical simulations were used to investigate a dust particle’s behavior in a Microparticle Electrodynamic Ion Trap (MEIT) with quadrupole geometry. Regions of stable confinement of a single particle are reported, in dependence of frequency and charge-to-mass ratio. An increase of the medium’s dynamical viscosity results in extending the confinement region for charged particles. Ordered Coulomb structures of charged dust particles obtained in the quadrupole trap operated in air, at atmospheric pressure, are also reported by the authors. Very recent papers report on measuring the charge of a single dust particle \cite{Lapitsky2016, Deputatova2015a}, on the effective forces that act on a microparticle confined in a Paul trap \cite{Lapitsky2015b}, or trapping of microparticles in gas flow \cite{Lapitsky2015a}. Another paper reports micrometer sized particles dynamic confined in a linear electrodynamic trap, at normal temperature and pressure conditions, where time variations of the light intensity scattered by trapped microparticles were recorded and analyzed in the frequency domain \cite{Visan2013}.

A recent paper on electrodynamic traps and the physics associated to them \cite{Libbrecht2015} studies ring electrode geometry and quadrupole linear trap geometry centimeter sized traps, as tools for physics teaching labs and lecture demonstrations. The authors introduce a viscous damping force to characterize the motion of particles confined in MEITs operating in air. For the microparticle species used, the authors show that Stokes damping describes well the mechanism of particle damping in air. The secular force in the one-dimensional case is inferred. Nonlinear dynamics is also observed, as such mesoscopic systems are excellent tools to perform integrability studies and investigate quantum chaos. 

Investigations on higher pole traps intended to be used for frequency standards based on trapped ions were first reported by Prestage and his group \cite{Prestage1999}, that built a 12-pole trap and have shuttled ions into it from a linear quadrupole trap. The outcome of the experiments was a clear demonstration on trapping larger ion clouds with respect to a quadrupole trap. The paper also emphasizes on the issue of space charge interactions that are non-negligible in a multi-pole trap. The Boltzmann equation describing large ion clouds in the general multipole trap of arbitrary order was solved. The authors state that fluctuations of the number of trapped ions influence the clock frequency much less severely than in the quadrupole case, which motivates present and future  quest towards developing and testing multipole traps for ion trap based high-precision frequency standards \cite{March2010, Knoop2014}. An interesting paper \cite{Burt2006} reports on the JPL multi-pole linear ion trap standard (LITS), which has demonstrated excellent frequency stability and improved immunity from two of its remaining systematic effects, the second-order Doppler shift and second-order Zeeman shift. The authors report developments that reduce the residual systematic effects to less than $ 6 \times 10^{−17} $, and the highest ratio of atomic transition frequency to frequency width (atomic line $Q$) ever demonstrated in a microwave atomic standard operating at room temperature.  

An Electron Spectrometer MultiPole Trap (ES-MPT) setup was devised by Jusko and co-workers \cite{Jusko2012}. A radio-frequency (RF) ion trap and an electron spectrometer were used in the experiment, with an aim to estimate the energy distribution of electrons produced within the trap. Results of simulations and first experimental tests with monoenergetic electrons from laser photodetachment of $O^{-}$ are presented. Other papers of the group from the Charles Univ. of Prague approach the study of associative photodetachment of $H^{-} + H$ using a 22-pole trap combined with an electron energy filter \cite{Roucka2009}, study of capture and cooling of $OH^{-}$ ions in multipole traps  \cite{Trippel2006, Otto2009} or $H^{-}$ ions in rf octopole traps with superimposed magnetic field \cite{Roucka2010}.

Multipole ion traps designs based on a set of planar, annular, concentric electrodes are presented in \cite{Clark2013}. Such mm scale traps are shown to exhibit trap depths as high as tens of electron volts. Several example traps were investigated, as well as scaling of the intrinsic trap characteristics with voltage, frequency, and trap scale. Stability and dynamics of ion rings in linear multipole traps as a function of the number of poles were investigated in \cite{Cartarius2013}. Multipole traps present a flatter potential in their centre and therefore a modified density distribution compared to quadrupole traps. Crystallization processes in multipole traps are investigated in \cite{Champenois2013}, where the dynamics and thermodynamics of large ion clouds in traps of different geometry is studied. Applications of these traps span areas such as QIP, metrology of frequencies and fundamental constants \cite{Quint2014, Knoop2015, Orszag2016}, production of cold molecules and the study of chemical dynamics at ultralow temperatures (cold ion–atom collisions).

\subsection{Multipole trap geometries and electronic setup}

\begin{figure}[bth]
    \begin{center}
    \includegraphics[scale=0.85]{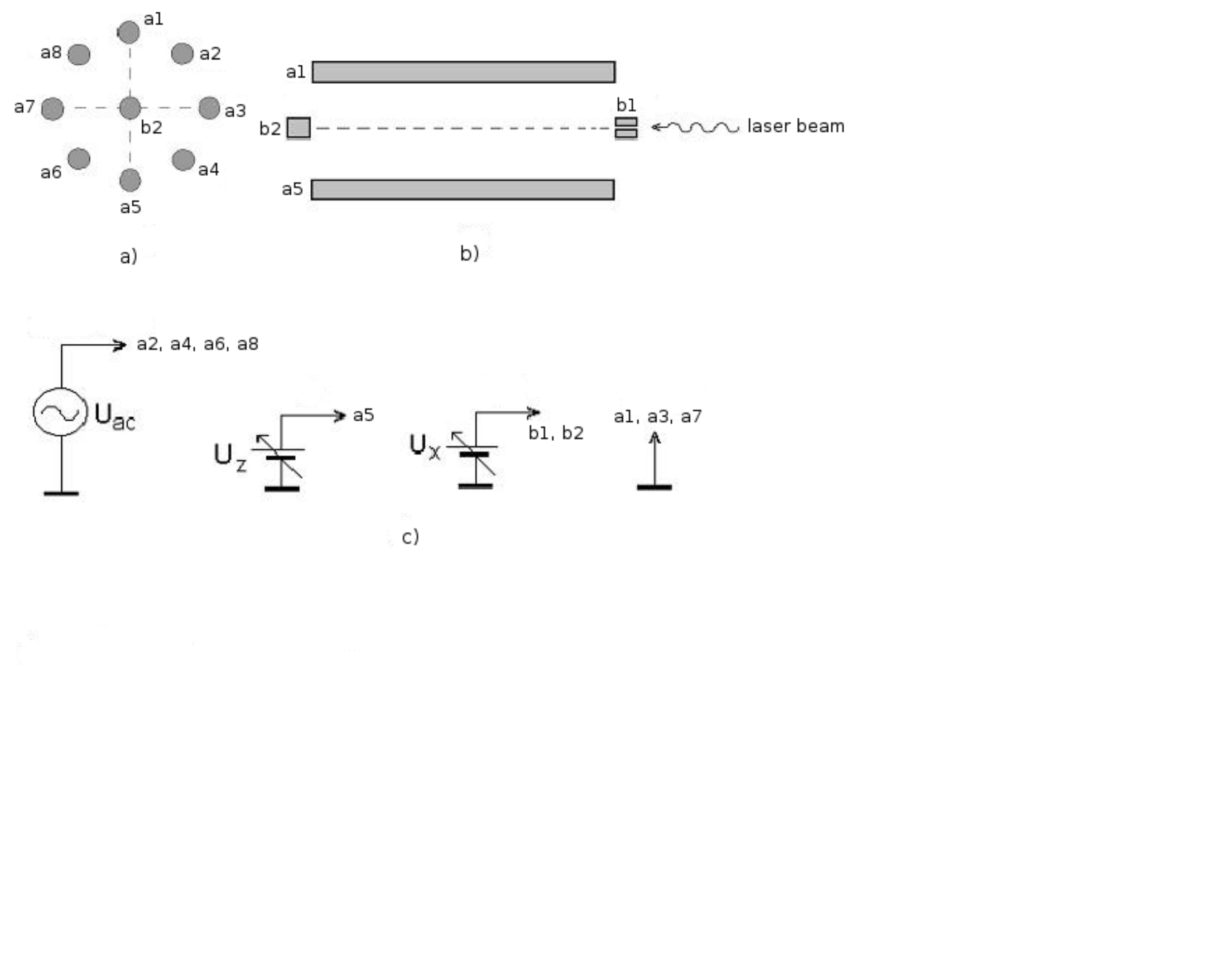} 
    \end{center}
    \vspace{-7.5cm}
    \caption{A sketch of the 8-electrode linear Paul trap geometry. a) cross-section;  
    \newline b) longitudinal section; c) electrode wiring}
    \label{octupole}
\end{figure}

We report two simple setups of Multipole Microparticle Electrodynamic Ion Traps (MMEITs). The first geometry consists of eight brass electrodes of 6 mm diameter, equidistantly spaced on a 20 mm radius, and two endcap electrodes located at the trap ends. The trap length is around 65 mm. The second trap geometry consists of twelve brass electrodes (a$_1$ - a$_{12}$) equidistantly spaced, and two endcap electrodes ($b_1$ and $b_2$). The electrode diameter is 4 mm, the trap radius is 20 mm, while the length of the electrodes does not exceed 85 mm. A sketch of the 8-electrode trap geometry we designed is shown in Fig. \ref{octupole}.

\begin{figure}[bth]
\begin{center}
\includegraphics[scale=0.85]{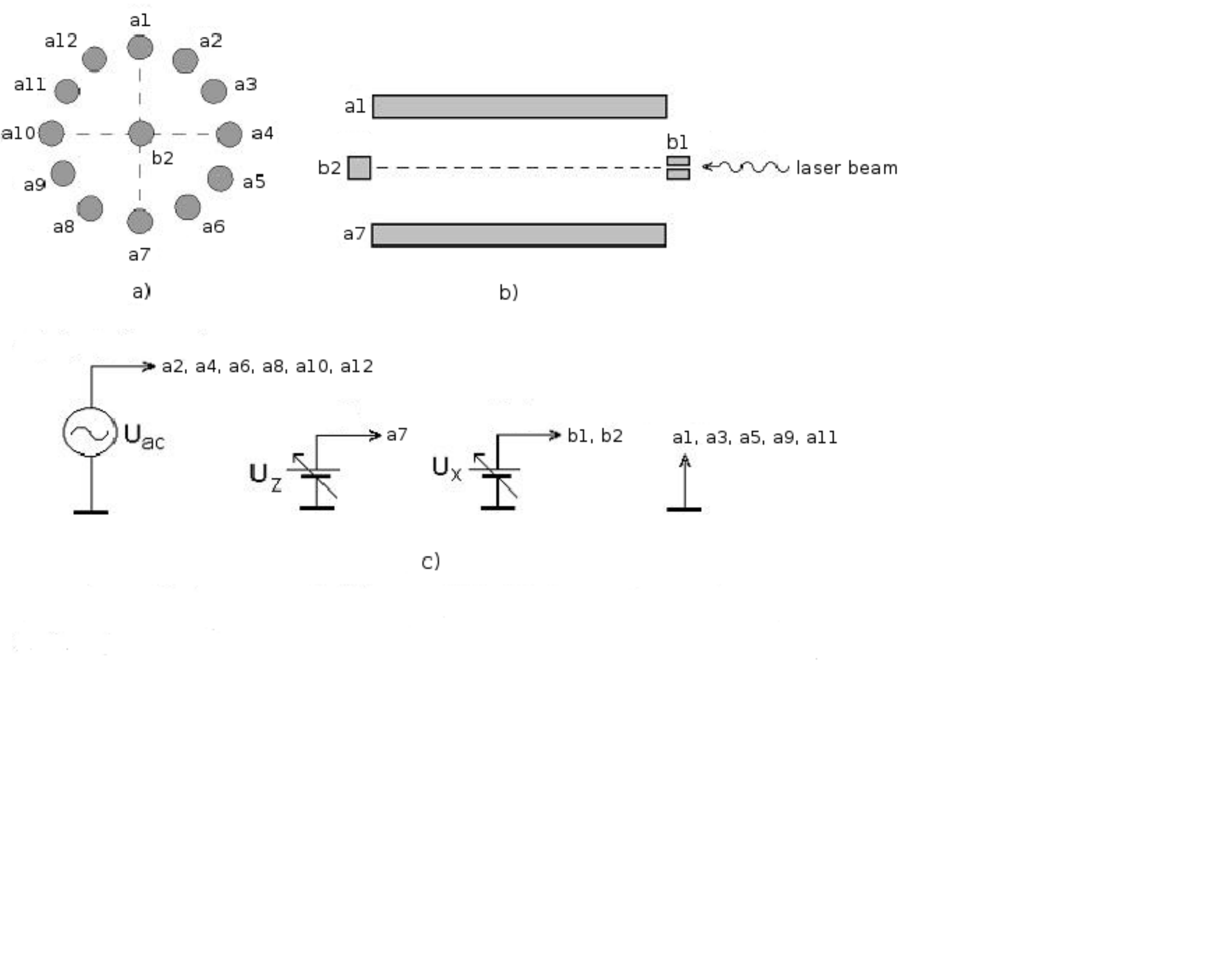} 
\end{center}
\vspace{-5cm}
\caption{A sketch of the 12-electrode linear Paul trap geometry. a) cross-section;  
\newline b) longitudinal section; c) electrode wiring }
\label{dodecapole}
\end{figure}

The 12-electrode (pole) trap geometry is shown in Fig. \ref{dodecapole}. Both setups are intended for studying the appearance of stable and ordered patterns for different electrically charged microparticle species. Alumina microparticles (with dimensions ranging from 60 microns up to 200 microns) were used in order to illustrate the trapping phenomenon, but other species can be considered. Specific charge measurements over the trapped microparticle species are expected to result, as the setup will be refined. The 12-electrode Paul trap which we designed is characterized by a variable geometry. The $b_2$ endcap electrode is located on a piston which slides along the $x$ axis of the trap. Hence, the trap length can vary between 10 mm up to 75 mm.  

\begin{figure}[bth]
\begin{center}
\includegraphics[scale=0.5]{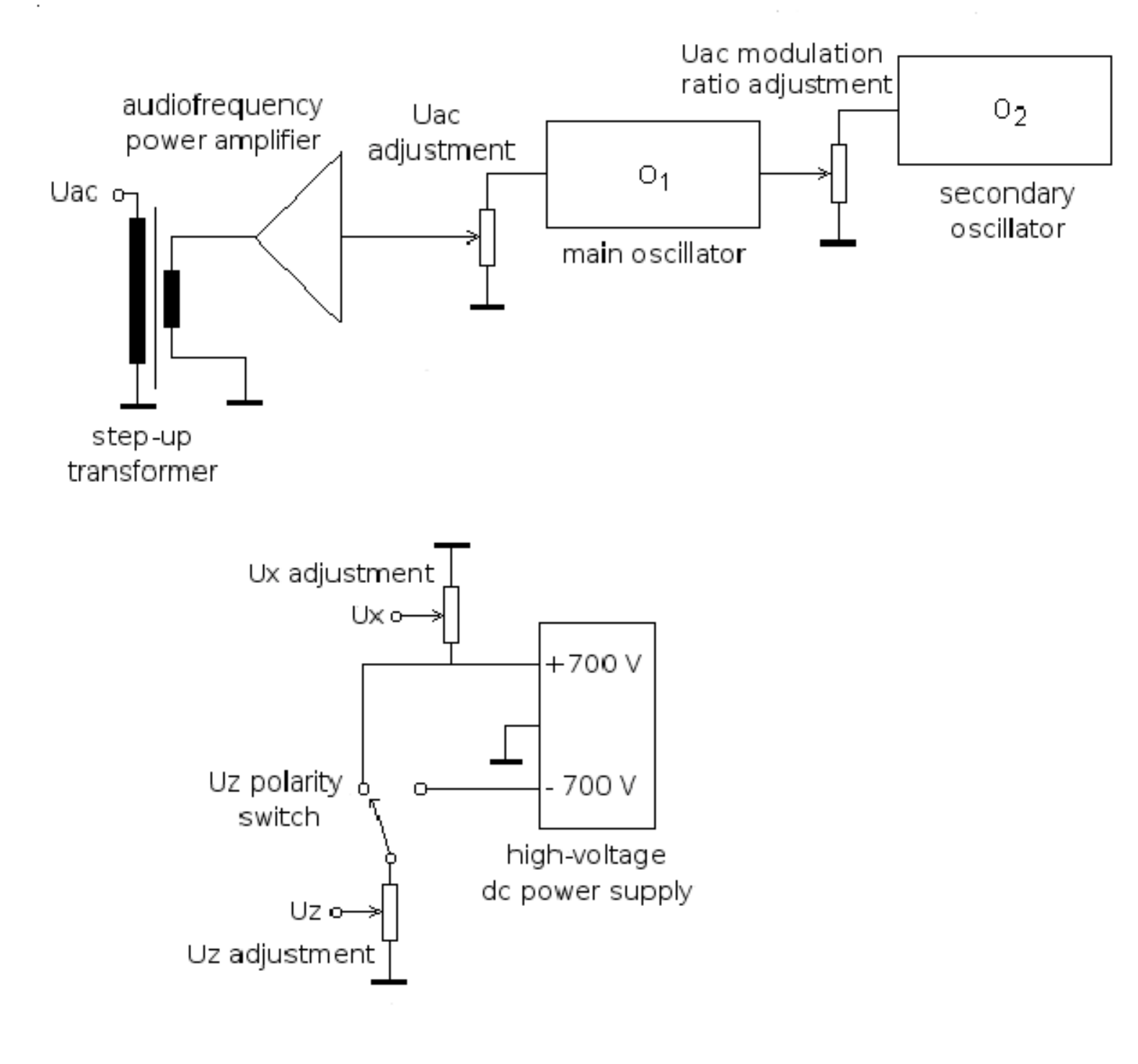}
\end{center}
\caption{Block scheme of the electronic supply unit}
\label{supply}  
\end{figure}

An electronic supply system was designed and realized. It supplies the a.c. voltage $V_{ac}$, with an amplitude of $0-4$ kV and a variable frequency in the $40-800$ Hz range, required in order to achieve radial trapping of charged particles. A high voltage step-up transformer, driven by a low frequency oscillator (main oscillator) $O_1$, delivers the $V_{ac}$ voltage as shown in Fig. \ref{supply}. An auxiliary oscillator $O_2$ is used to modulate the amplitude of the $V_{ac}$ voltage. The modulation ratio can reach a peak value of $100\%$, while the modulation frequency lies in the $10-30$ Hz range. 

\begin{figure}[bth]
\begin{center}  
\includegraphics[scale=0.4]{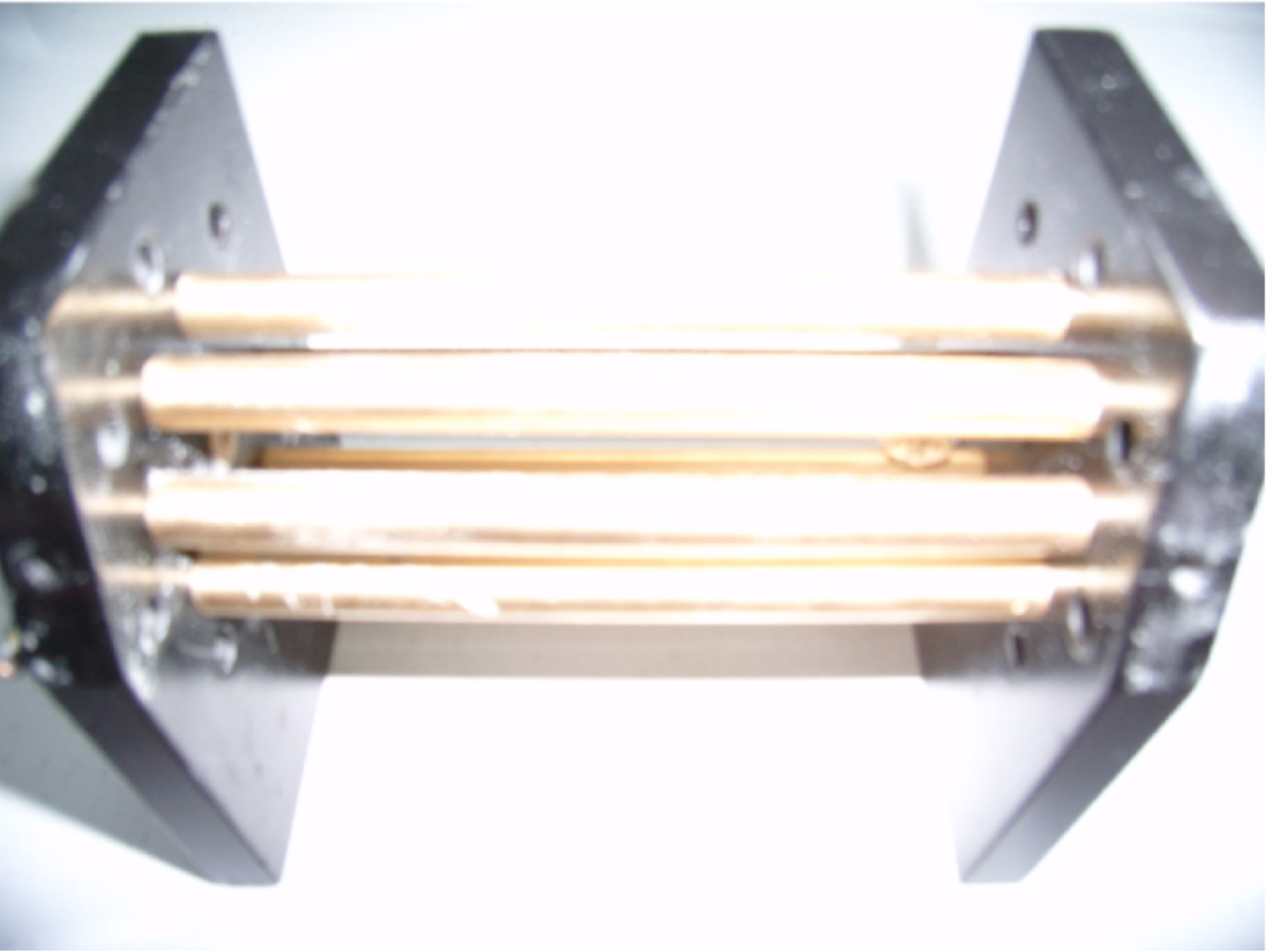}
\hspace{1cm}  
\includegraphics[scale=0.43]{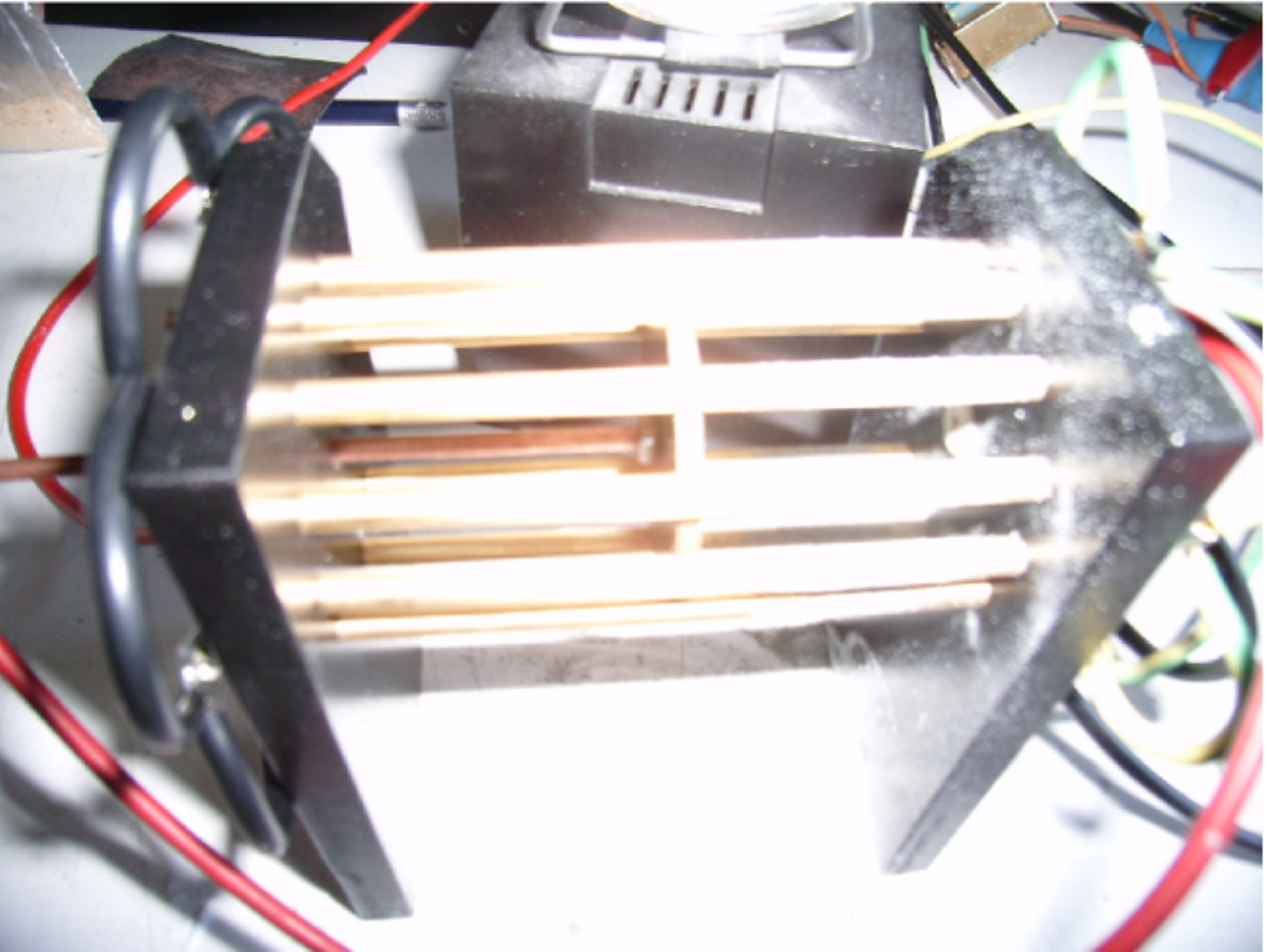} 
\end{center}
\caption{Photos of the 8-electrode and 12-electrode linear Paul trap geometries} 
\label{traps}
\end{figure}

Photos of both the 8-electrode and 12-electrode traps are presented in Fig. \ref{traps}. The electronic system also supplies a variable d.c voltage $U_z$ (also called diagnose voltage), applied between the upper and lower multipole trap electrodes, used to compensate the gravitational field and shift the particle position along the $z$ axis. Ions confined in such traps under ultrahigh vacuum conditions do arrange themselves along the longitudinal $x$ axis and within a large region around it, where the trapping potential is very weak. The situation is sensibly different in case of electrically charged microparticles, which we explain in Section \ref{model}. Another d.c. variable voltage $U_x$ is applied between the trap endcap electrodes, in order to achieve axial confinement and prevent particle loss near the trap ends. The $U_z$ and $U_x$ voltages have values ranging between $50-700$ V, while their polarity can be reversed. 

Both $U_z$ and $U_x$ voltages are obtained using a common d.c. double power supply, which delivers a voltage of $\pm 700$ V at its output. Because the absorbed d.c. current is very low, the d.c. voltages are supplied to the electrodes by means of potentiometer voltage dividers as shown in Fig. \ref{supply}. 

\begin{figure}[bth]
\begin{center}  
\includegraphics[scale=0.275]{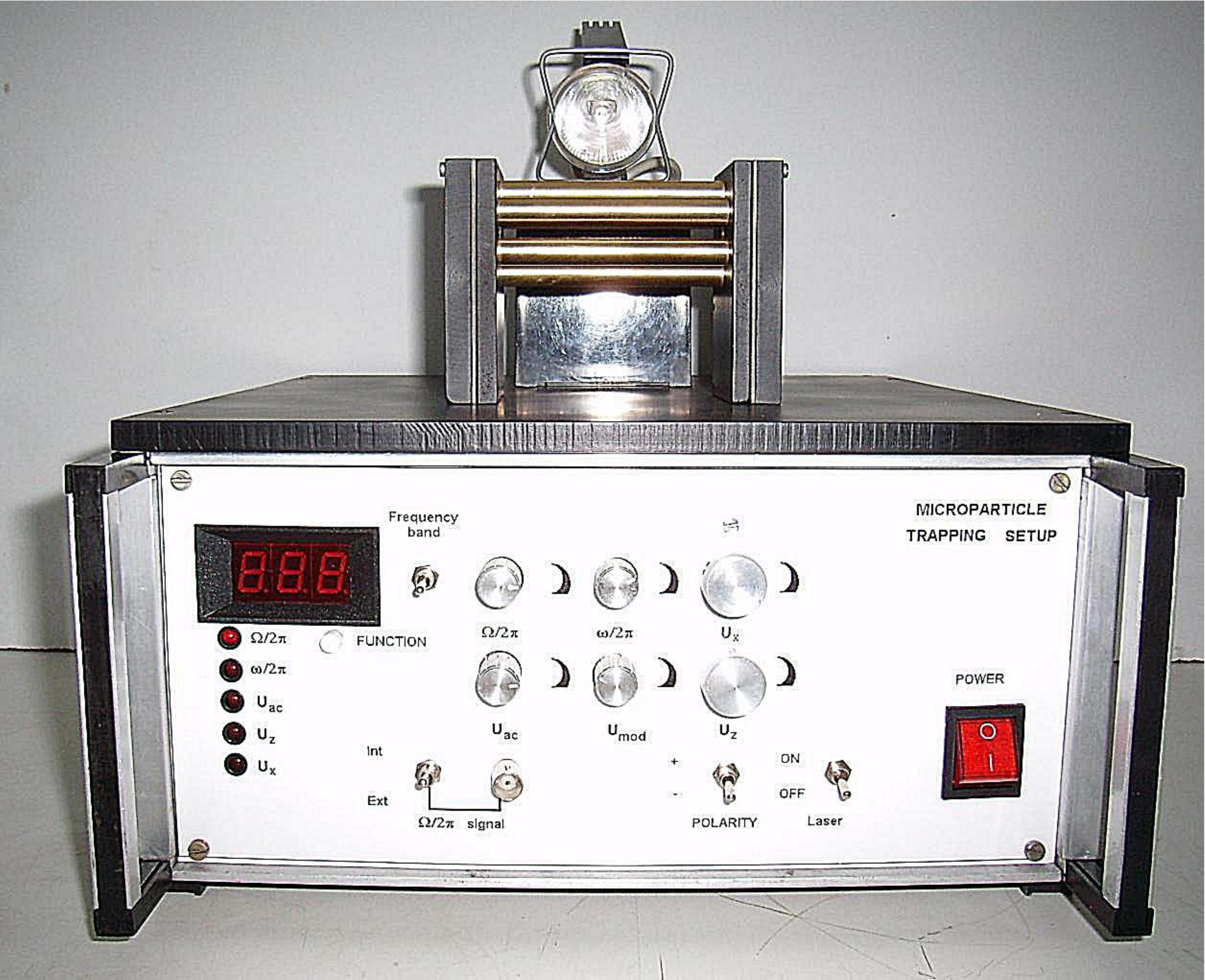}
\end{center}
\caption{Electronic supply unit used for the 8-pole and 12-pole traps} 
\label{ACDCsupply}
\end{figure}

The supply system is a single unit which delivers all the necessary trapping voltages, as shown in Fig. \ref{ACDCsupply}. A microcontroller based measurement electronic circuit allows separate monitoring and displaying of the supply voltage amplitudes and frequencies (for the a.c. voltage and modulation voltage). In order to visualize and diagnose the trapped particles, the experimental setups have been equipped with two different illumination systems. The first system is based on a halogen lamp whose beam is directed normal to the trap axis. The second system consists of a laser diode whose beam is directed parallel to the trap axis, as one of the endcap electrodes is pierced.  

If the a.c. potential is not too high (usually less than 3 kV), stable oscillation will occur until the d.c. potential is adjusted to balance vertical forces such as gravity. When such condition is achieved, the oscillation amplitude becomes vanishingly small and the particle experiences stable trapping \cite{Kulkarni2011}. 
The traps under test are fitted to study complex Coulomb systems (microplasmas) confined in multipole dynamic traps operating in air, at SATP conditions. The paper brings new evidence on microparticle trapping, while it demonstrates that the stability region for multipole traps is larger with respect to a quadrupole trap (electrodynamic balance configuration) \cite{Prestage1999, Vasilyak2013}. 

\begin{figure}[bth]
\begin{center}  
\includegraphics[scale=0.5]{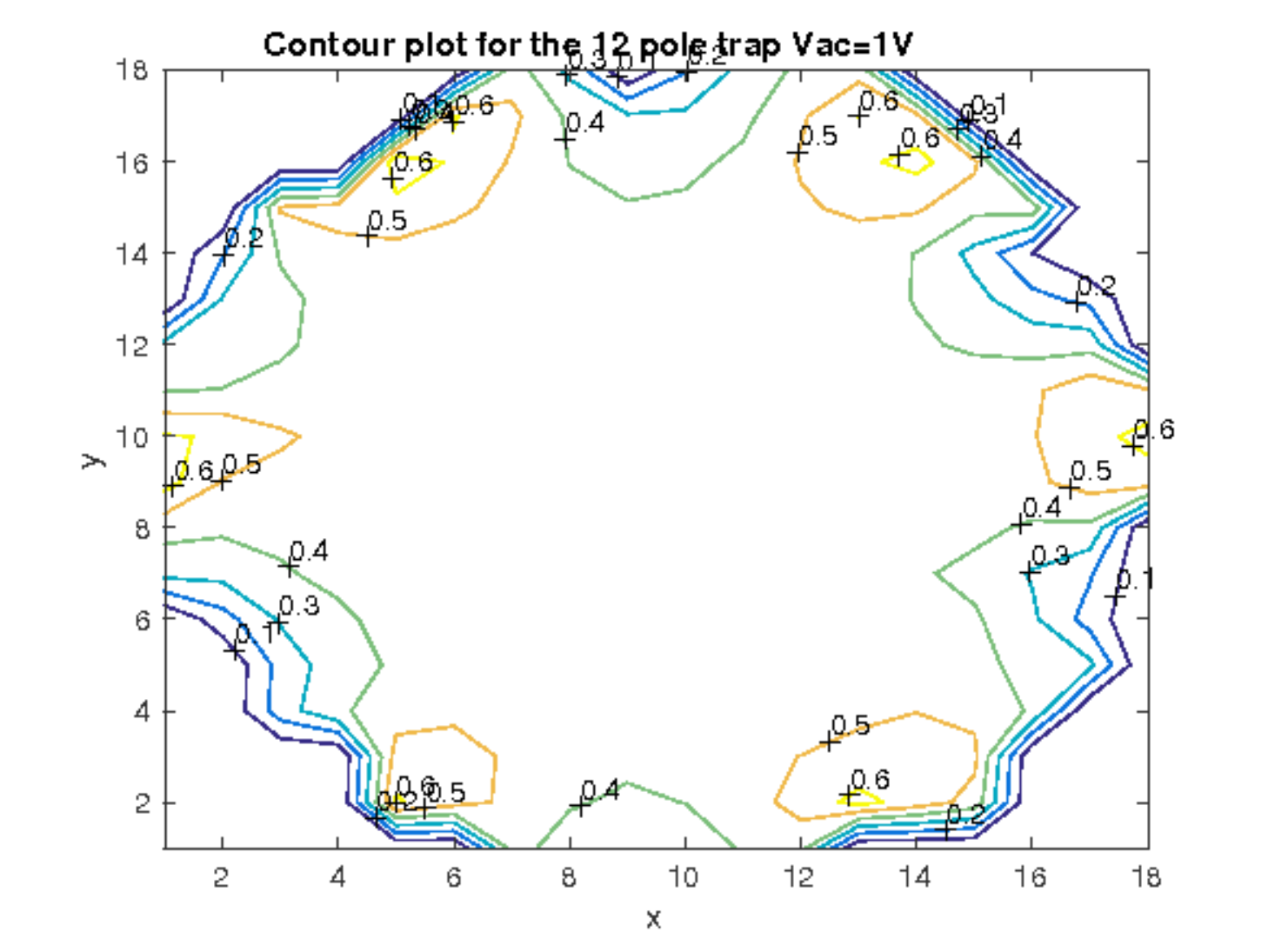}
\includegraphics[scale=0.5]{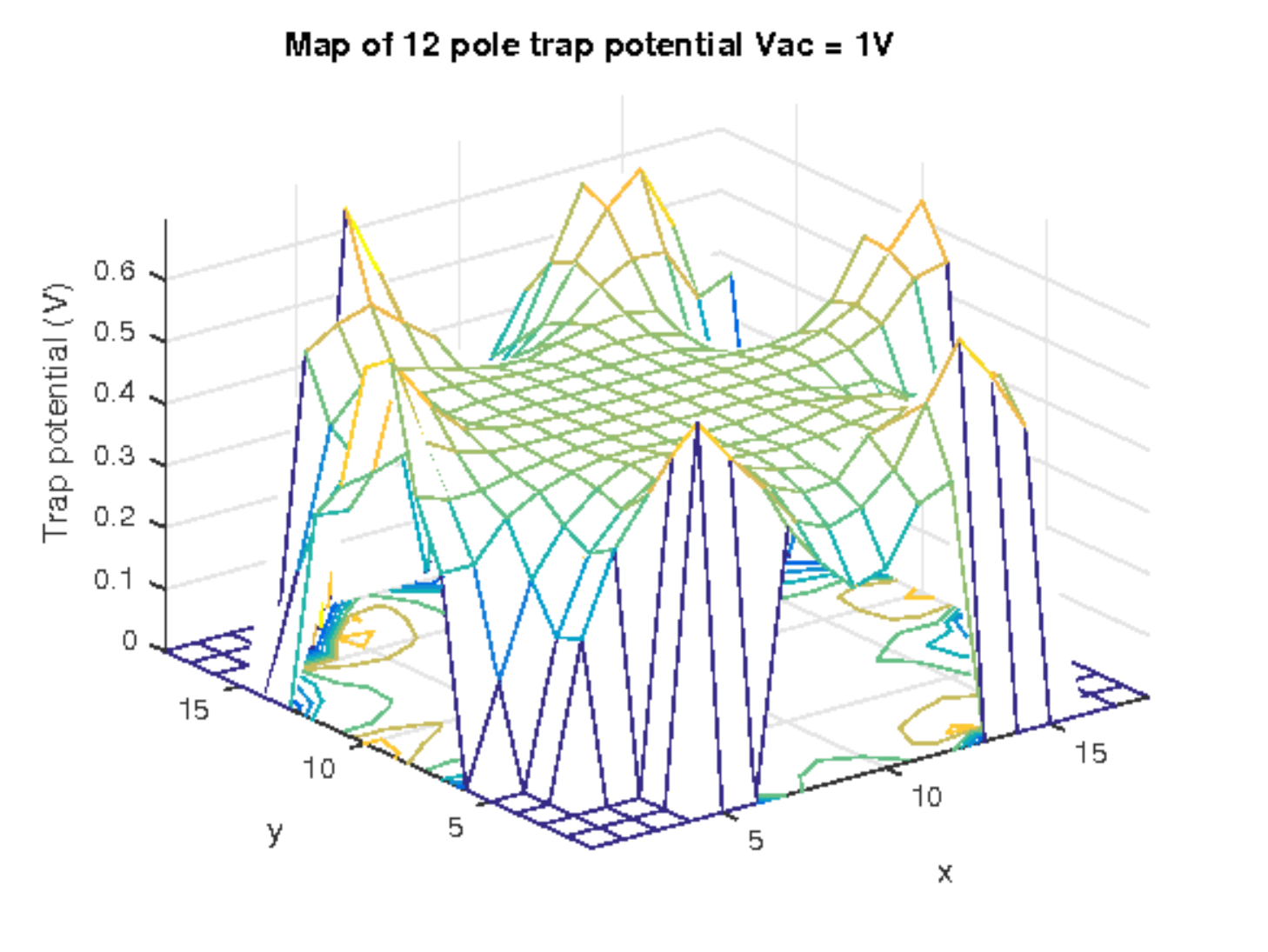}
\vspace{.5cm}
\includegraphics[scale=0.55]{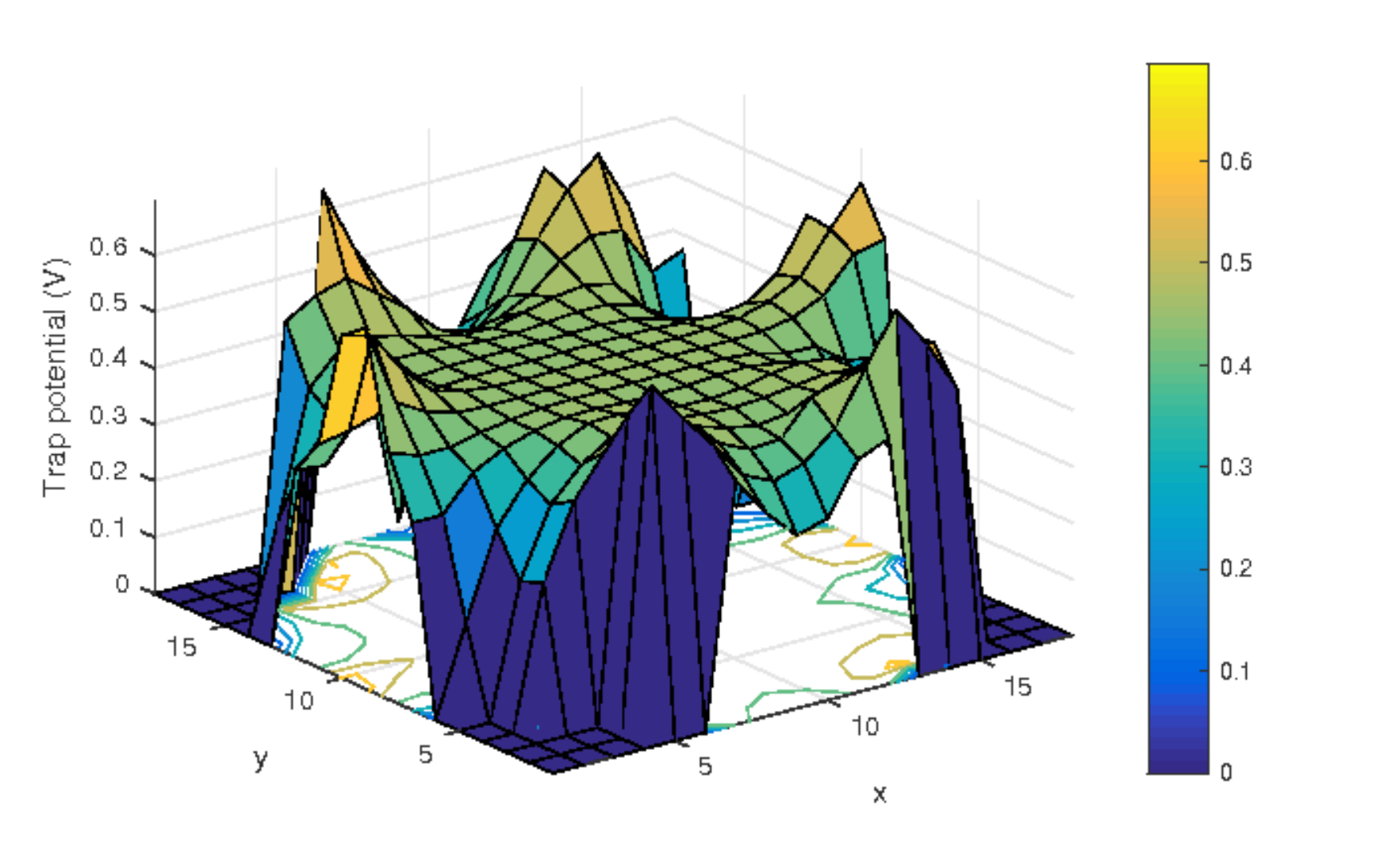}
\end{center}
\caption{Contour and 3D plots for the 12-electrode Paul trap potential when $V_{ac}=1 V$} 
\label{12pole1V}
\end{figure}

\begin{figure}[bth]
\begin{center}  
\includegraphics[scale=0.5]{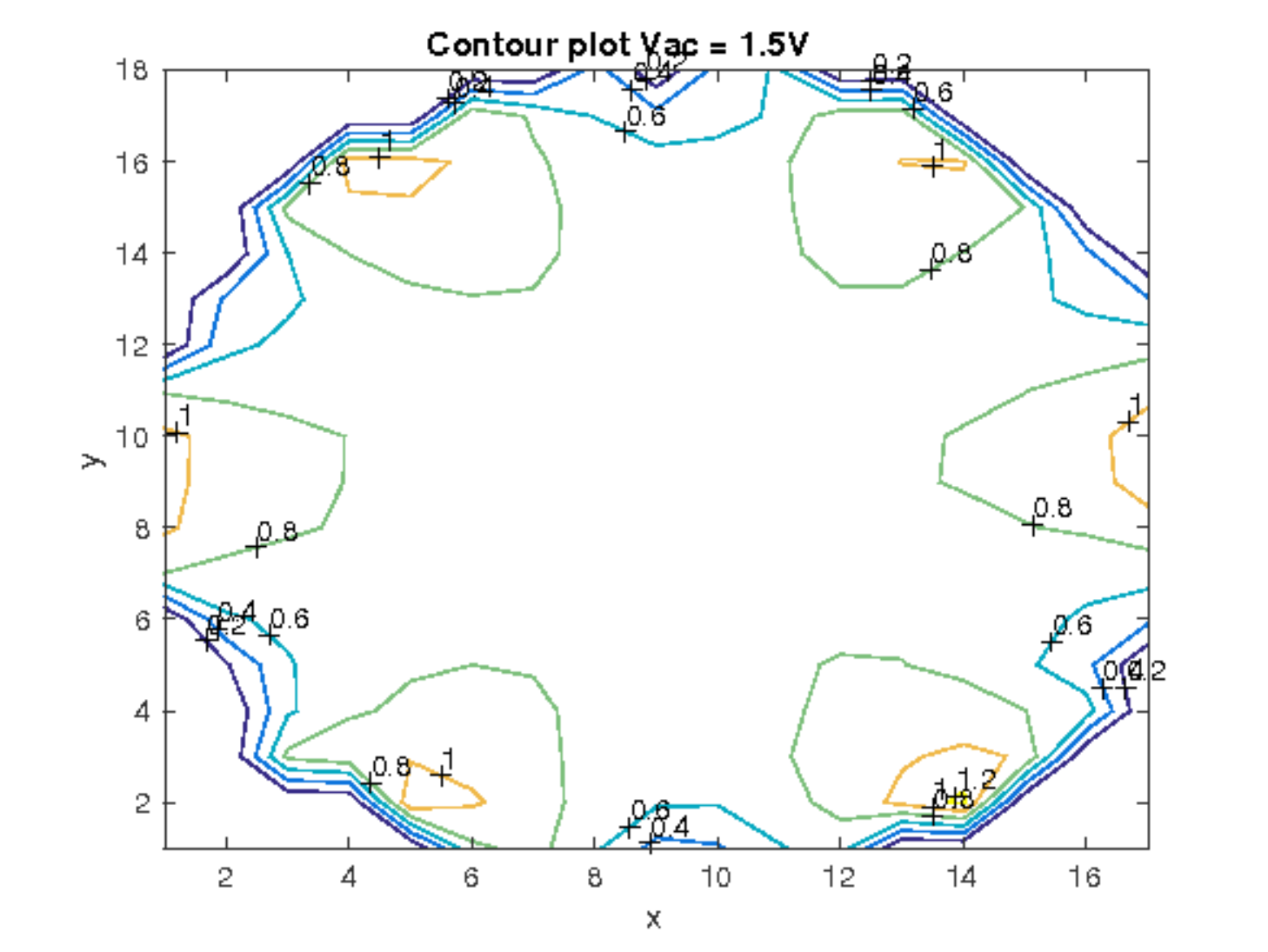}
\includegraphics[scale=0.5]{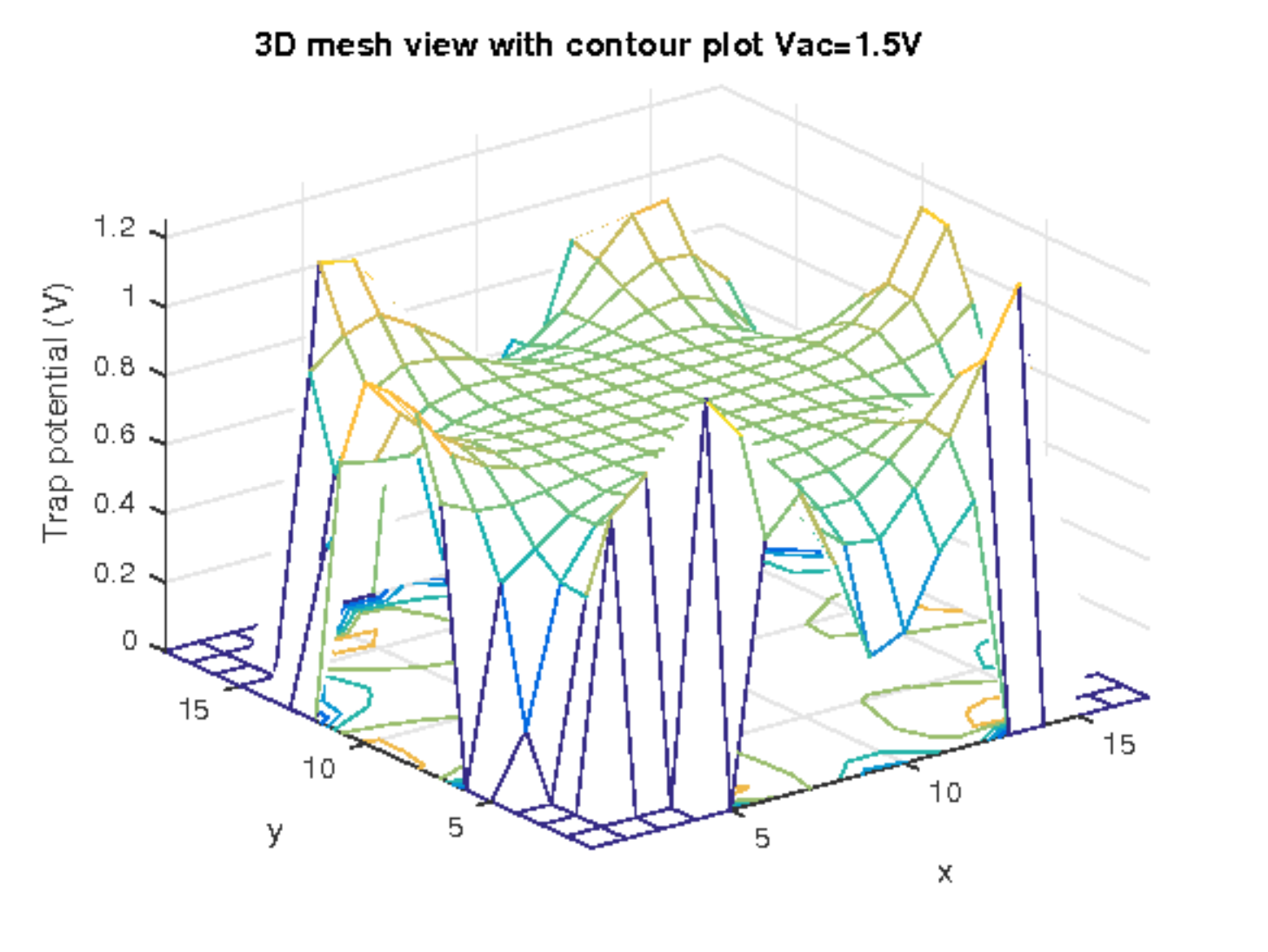}
\vspace{.5cm}
\includegraphics[scale=0.55]{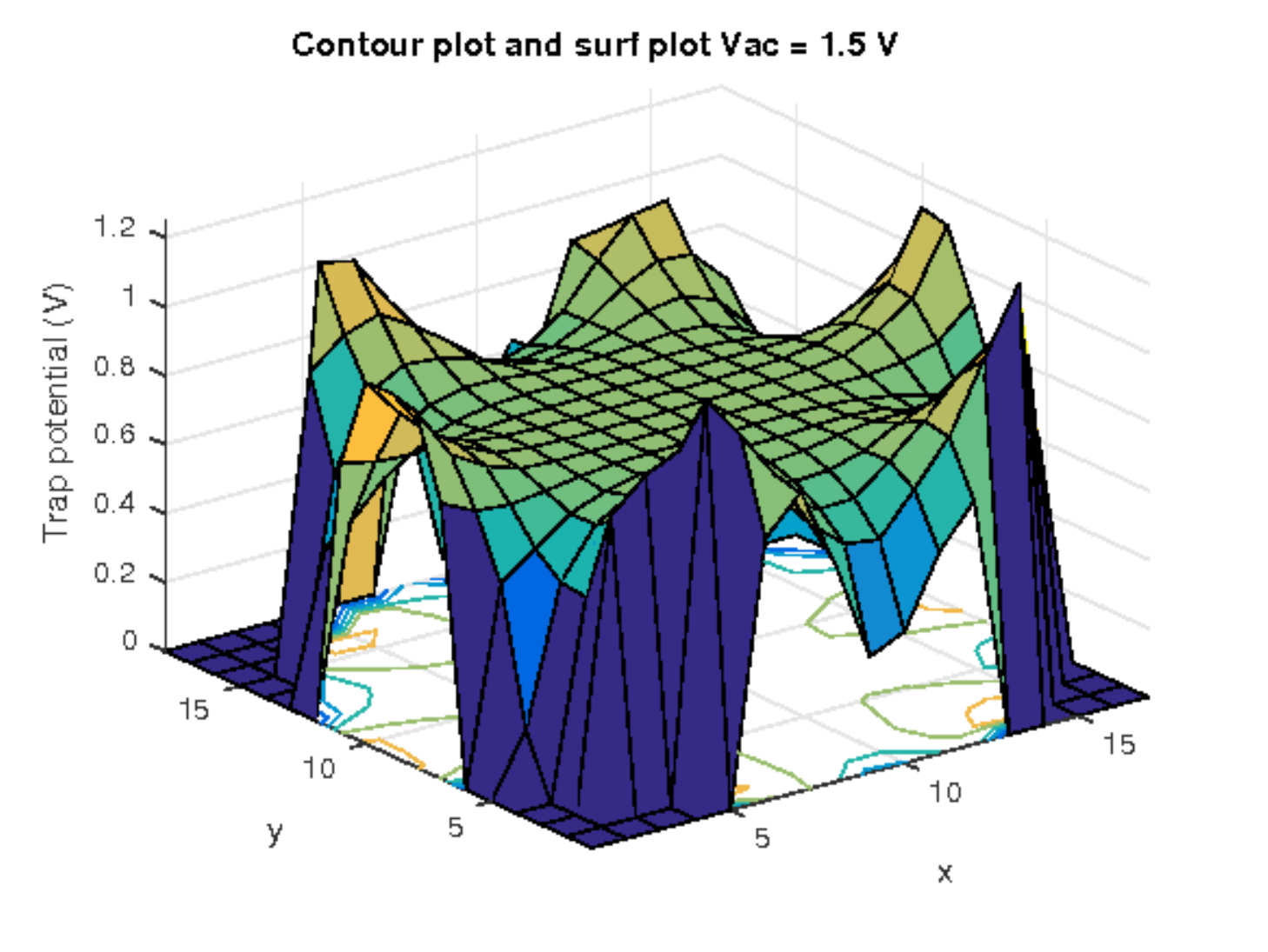}
\end{center}
\caption{Contour and 3D plots for the 12-electrode Paul trap potential when $V_{ac}=1.5 V$} 
\label{12pole15V}
\end{figure}

We have used the electrolytic tank method to map the radiofrequency field potential within the 12-electrode trap volume. We have designed and realized a precision mechanical setup that has enabled us to chart the field within the trap. The electrolyte solution used was distilled (dedurized) water. Two a.c. supply voltage values were used: 1 V and 1.5 V, respectively. We emphasize that these represent the amplitude values measured. Measurements were performed for an a.c. frequency value of $\Omega = 2\pi \times 10$ Hz. Figures \ref{12pole1V} and \ref{12pole15V} show the contour and 3 D maps of the trap potential (rms values) that we have obtained. 

\section{Results}\label{results}

We have investigated different linear multipole electrodynamic trap geometries operating in air, under SATP conditions. We have focused on an 8-electrode and a 12-electrode trap geometry respectively, intended for charged microparticle confinement and for illustrating the appearance of planar and volume structures for these microplasmas. Microparticles are radially confined due to the a.c. trapping voltage $V_{ac}$, which we set at a value of 2.5 kV. The microplasmas can be shifted both axially and vertically, using two d.c. voltages: $U_x$ and $U_z$. The geometry of the 8-electrode trap has proven to be very critical, different radii have been tested and the trap has gone through intensive tests with an aim to optimize it. It presents a sensibly higher degree of instability compared to the 12-electrode geometry. 

The 12-electrode (pole) trap has been studied more intensively as particle dynamics is more stable for such geometry. We have loaded the traps with microparticles by means of a miniature screwdriver. The peak of the screwdriver is inserted into the alumina powder. When touching the screwdriver to one of the trap electrodes, the particles are instantly charged and a small part of them are confined, depending on their energy and phase of the a.c. trapping field. Thus a trapped particle microplasma results (very similar to a dusty plasma, which is of great interest for astrophysics), consisting of tens up to hundreds of particles. Such a setup would be suited in order to study and illustrate particle dynamics in electromagnetic fields, as well as the appearance of ordered structures, crystal like formations. 
	
Practically, stable confinement has been achieved especially in the 12-electrode (pole) trap, as we have observed thread-like formations (strings) and especially 2D (some of them zig-zag) and 3D structures of microplasmas. The stable structures observed have the tendency of aligning with respect to the $z$ component of the radial field. The ordered formations were not located along the trap axis, but rather in the vicinity of the electrodes. The laser diode has been shifted away from the initial position (along the trap axis), towards outer regions of the trap. We report stable confinement for hours and even days. Laminary air flows within the trap volume break an equilibrium which might be described as somehow fragile and some of the particles can get lost at the electrodes. Nevertheless, we have also tested the trap under conditions of intense air flows and we have observed that most of the particles remain trapped, even if they rearrange themselves after being subject to intense and repeated perturbations. When a transparent plastic box was used in order to shield the trap, a sensible increase in the dynamical stability of the particle motion has been achieved. 
	
\begin{figure}[bth]
\begin{center}  
\includegraphics[scale=0.48]{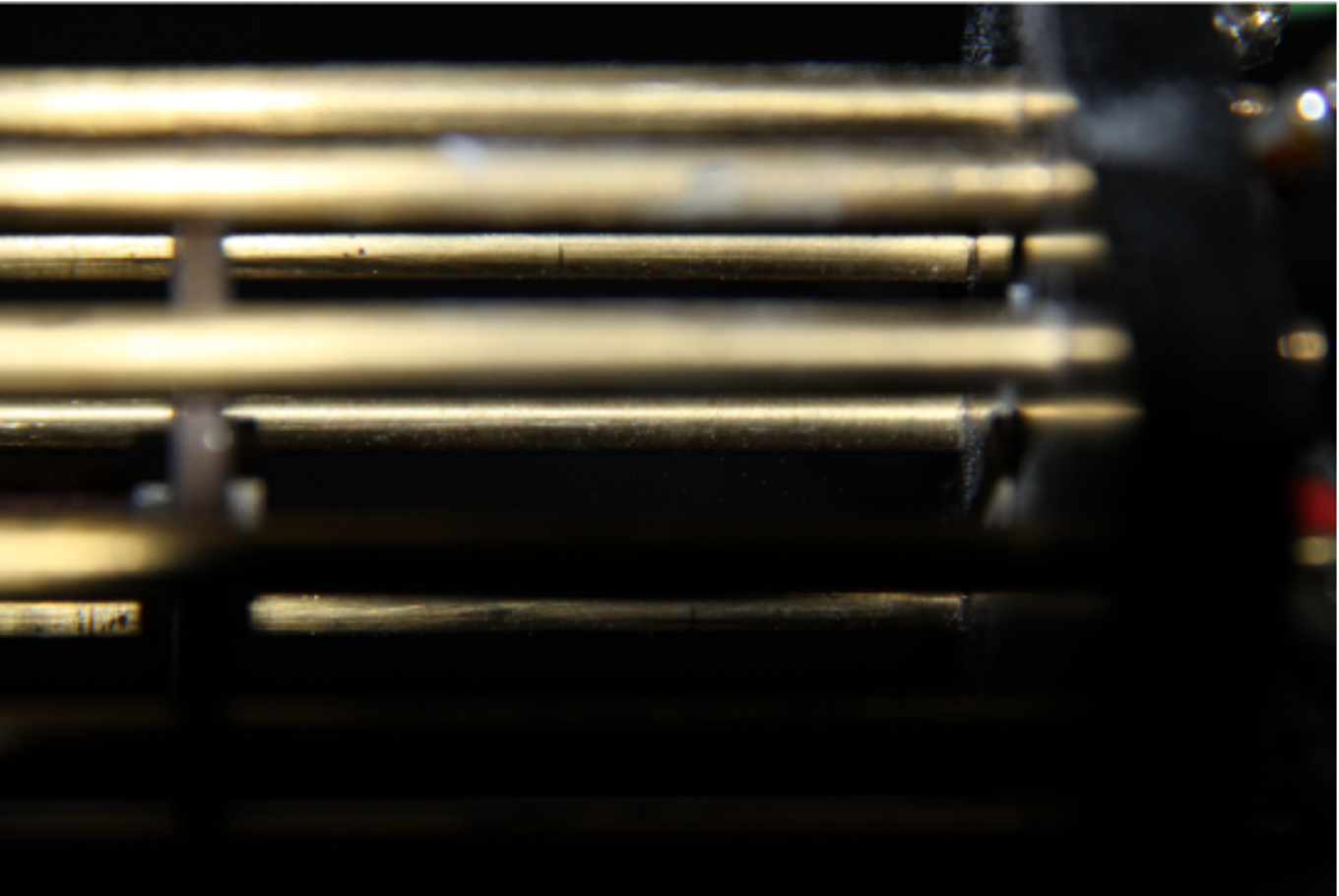}
\includegraphics[scale=0.43]{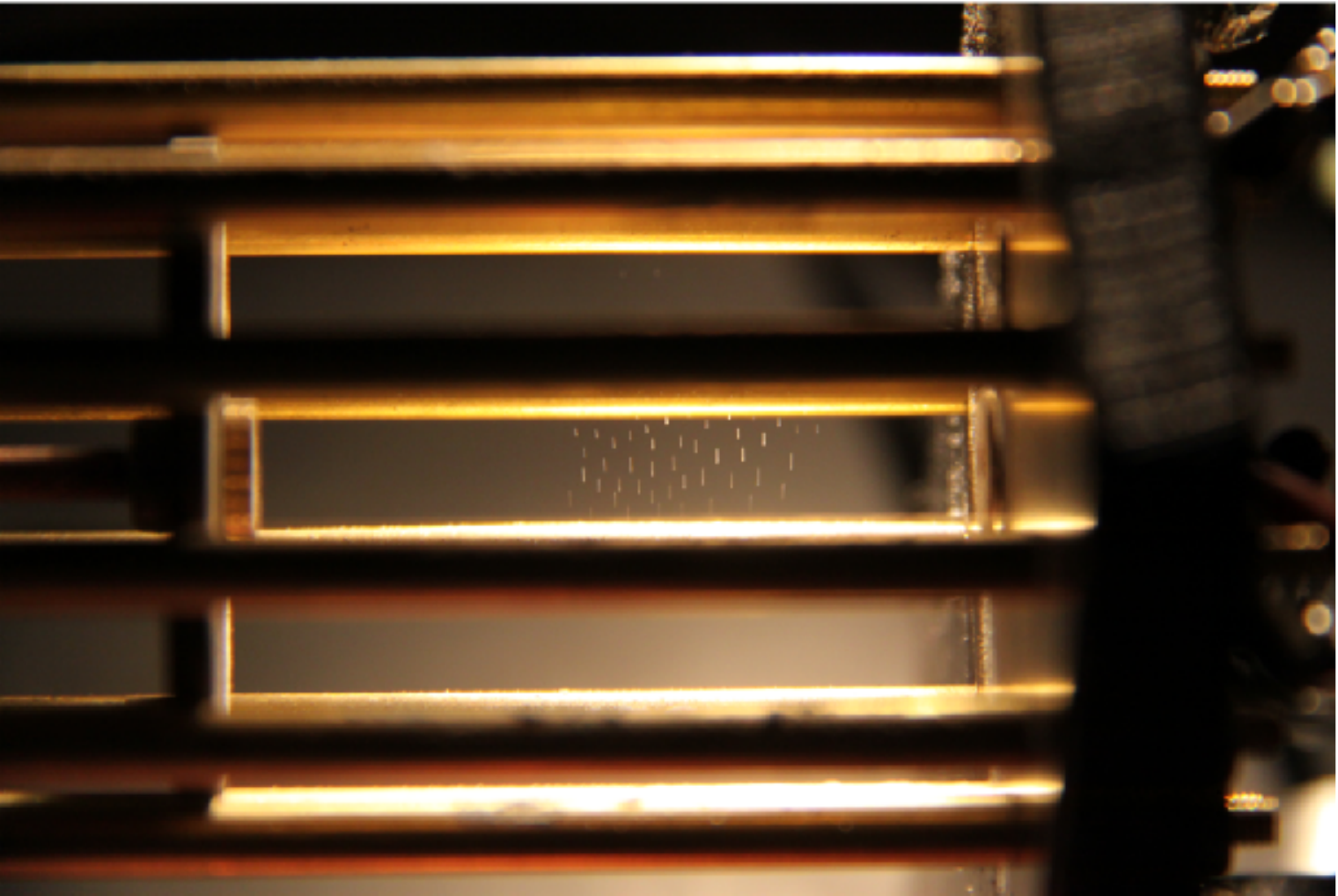}
\vspace{.5cm}
\includegraphics[scale=0.52]{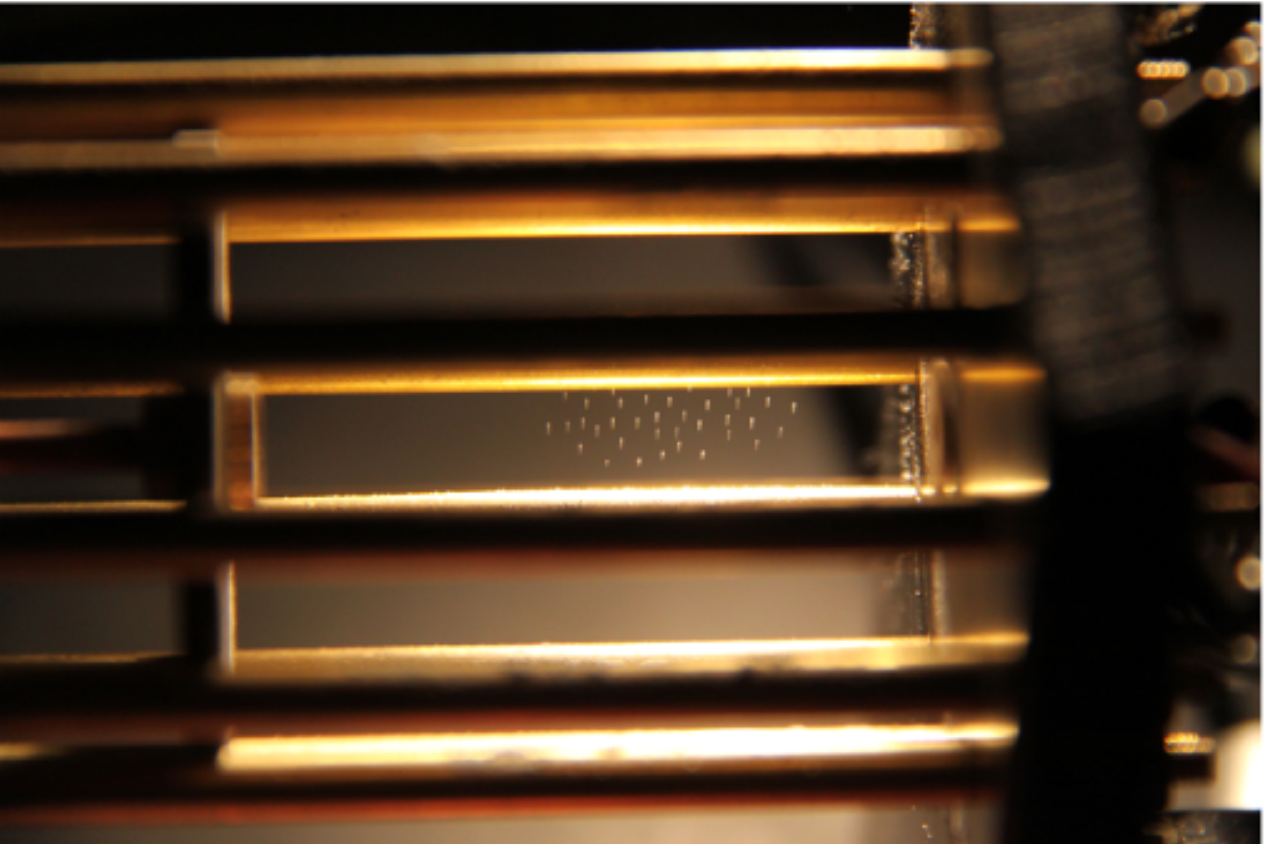}
\includegraphics[scale=0.51]{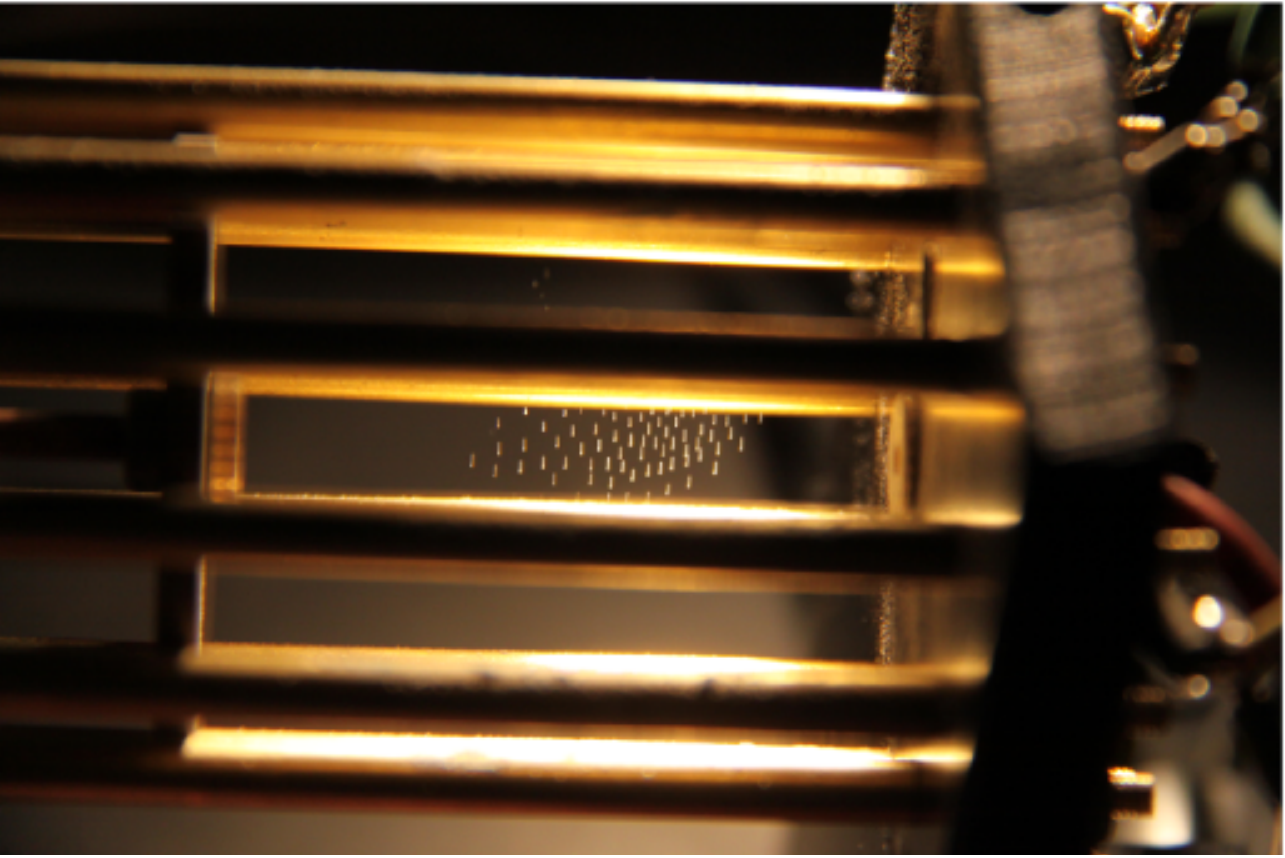}
\vspace{.5cm}
\includegraphics[scale=0.515]{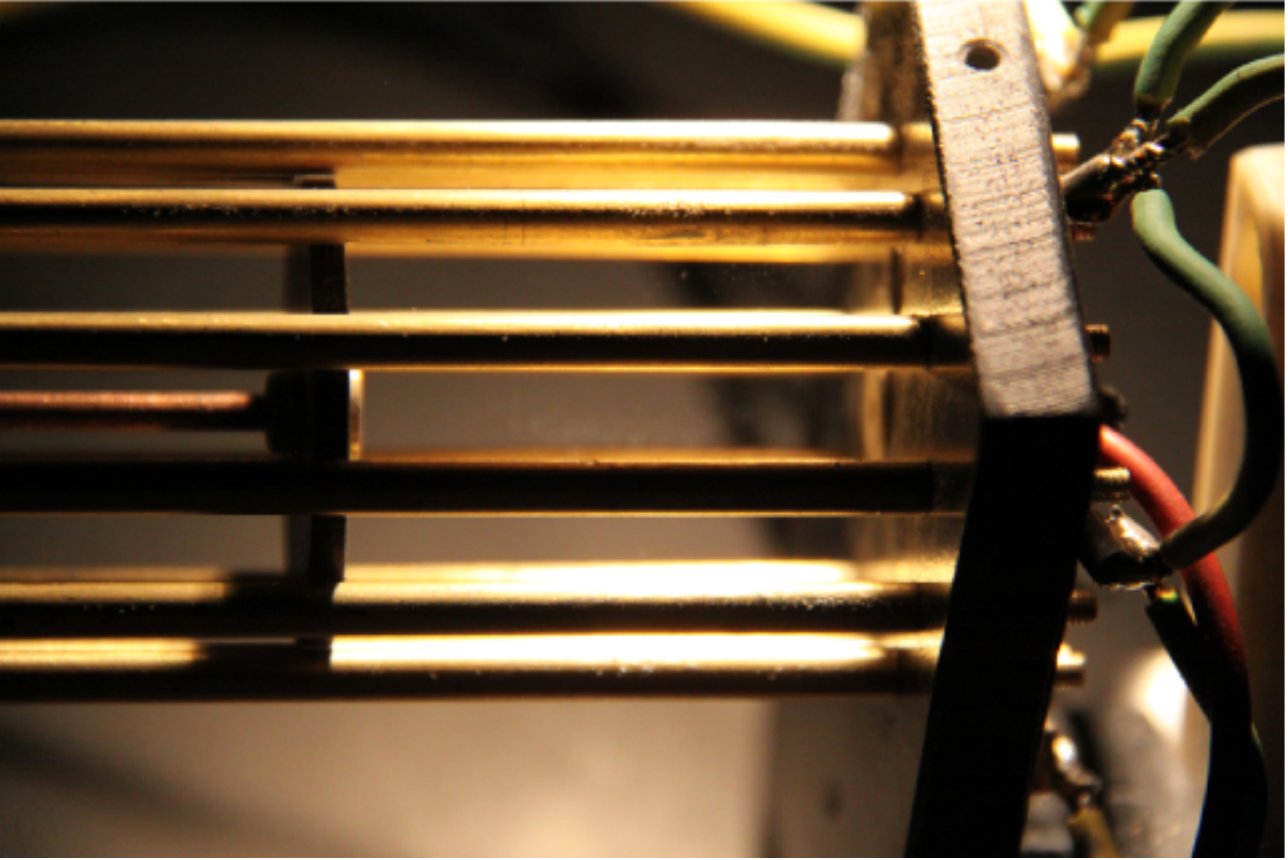}
\includegraphics[scale=0.49]{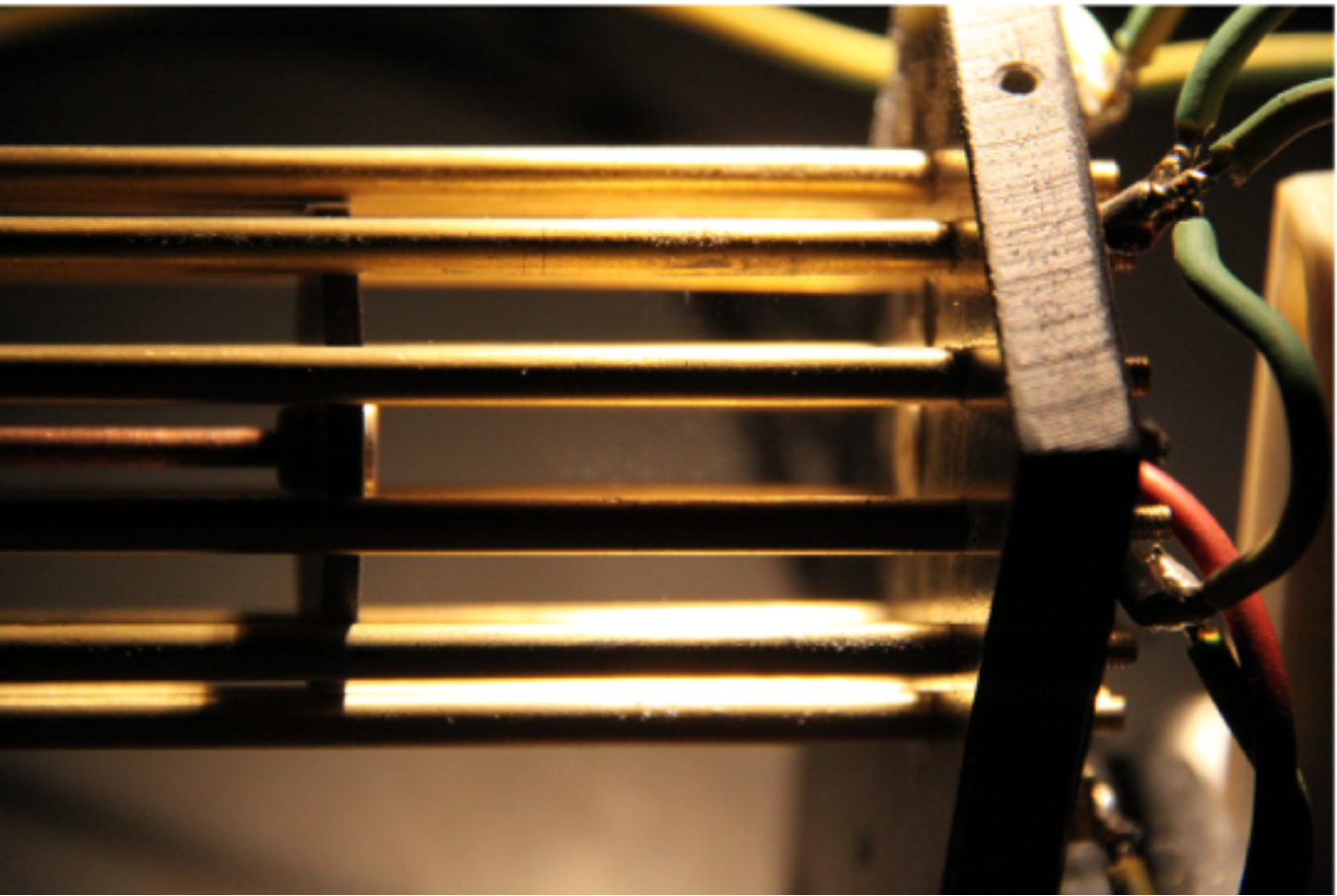}
\vspace{.5cm}
\includegraphics[scale=0.49]{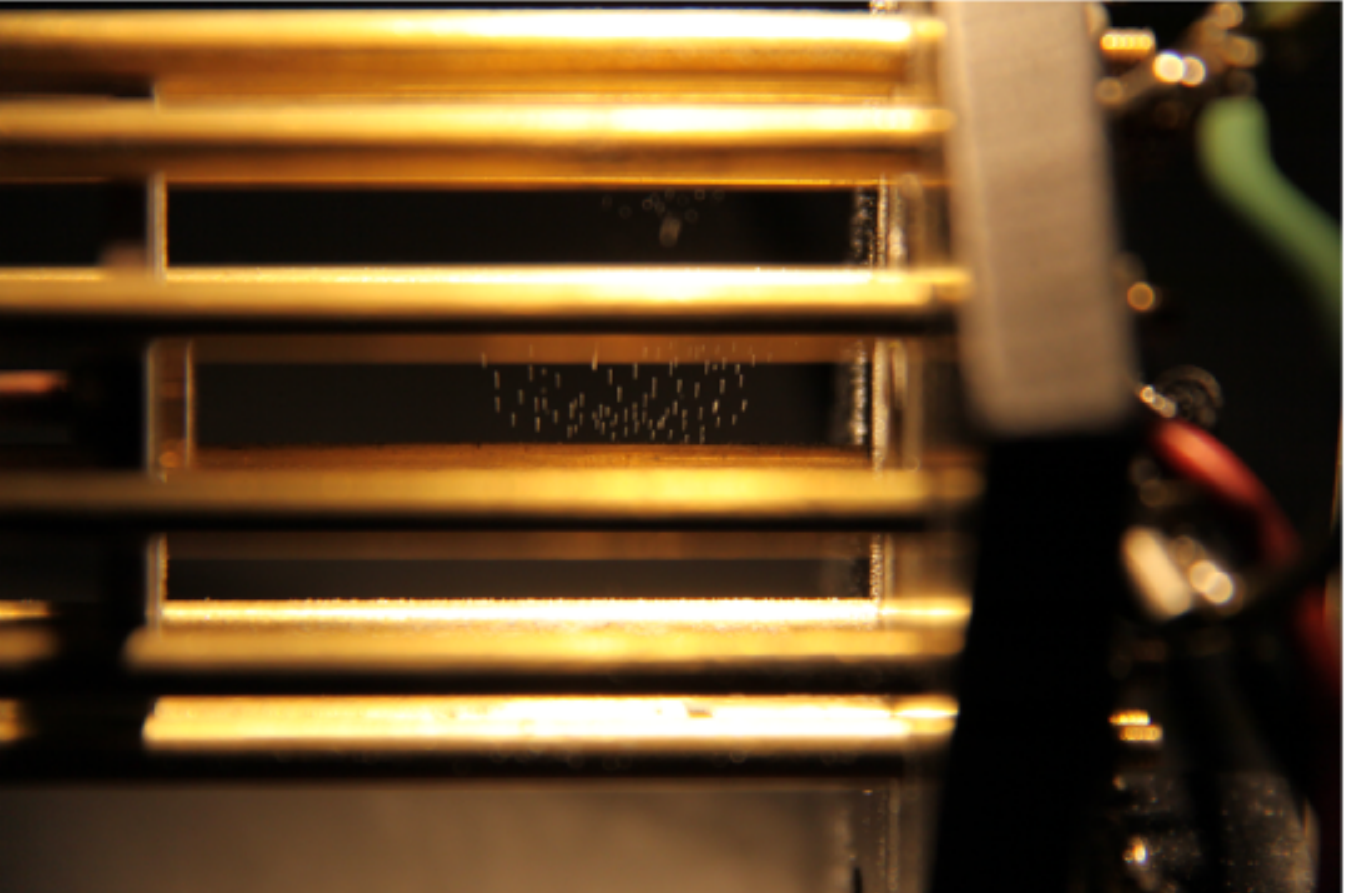}
\includegraphics[scale=0.5]{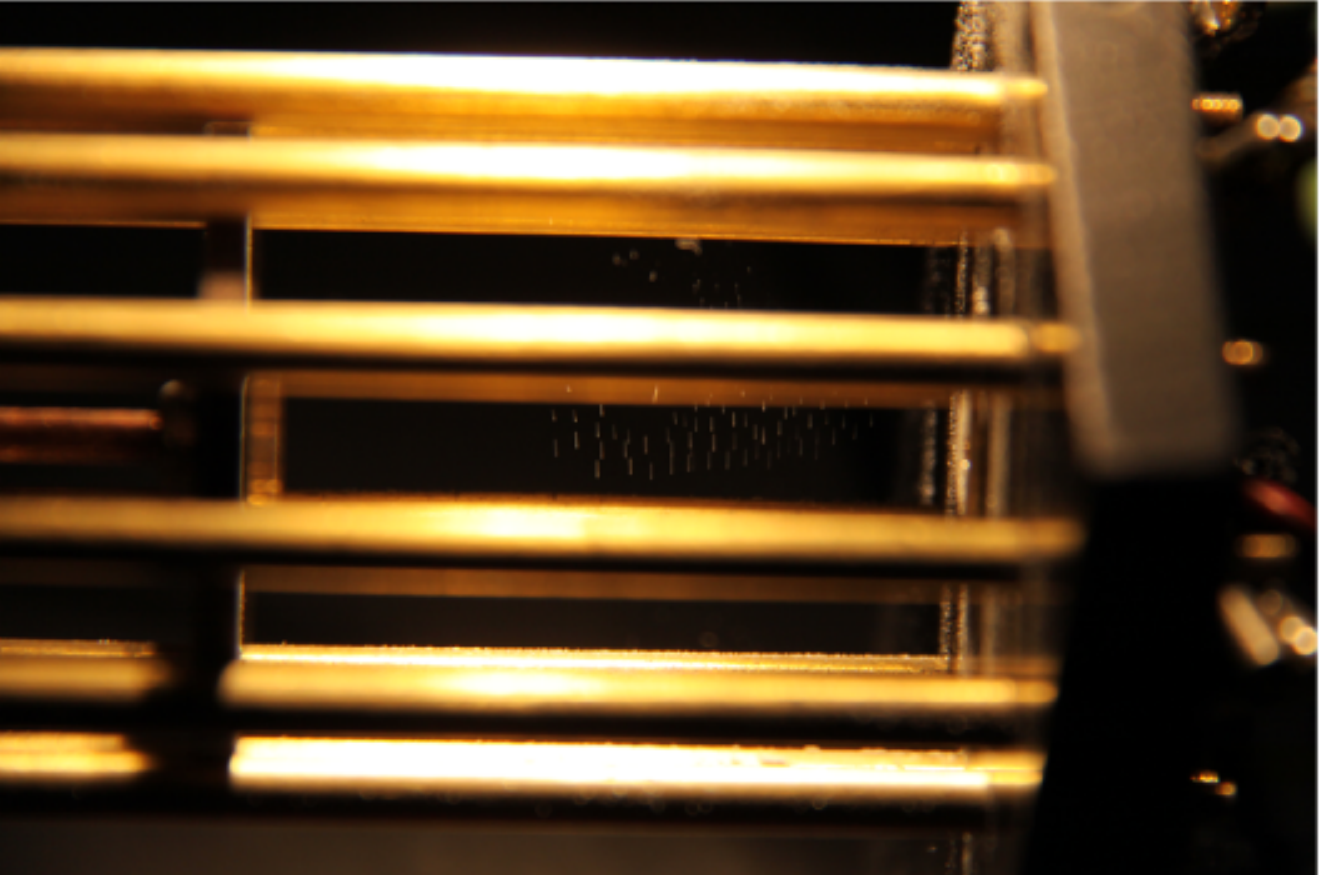}
\end{center}
\caption{Photos of the ordered structures observed in the 12-electrode Paul trap} 
\label{microplasmas}
\end{figure}

In Fig. \ref{microplasmas} we present a few pictures which illustrate the stable structures we have been able to observe and photograph. All photos were taken with a high sensitivity digital camera, using the halogen lamp in order to illuminate the trap. Pictures taken using the laser diode were less clear due to reflection of light on the trap electrodes and this is the reason why they are not included. We report filiform structures consisting of large number of microparticles far from the trap center, where the trapping potential is extremely weak. This leads to the conclusion that microparticle weight is not balanced by the trapping field in the region located near the trap center. Moreover, most of the ordered structures observed were generally located very close to the trap electrodes, at distances about 2-10 mm apart from them. The a.c. frequency range was swept between $\Omega = 2\pi \times 50$ Hz up to $\Omega = 2 \pi \times 100$ Hz. 

The trap geometries investigated are characterized by more regions of stable trapping, some of them located near the trap electrodes. Most of the photos taken illustrate such phenomenon along with the numerical simulations performed. At least three or four regions of stable trapping were identified, but due to camera limitations focusing was achieved for a limited region of space. When looking with bare eyes and for an adequate angle of observation, spatial structures were observed in different regions within the trap volume. We can ascertain that the multipolar trap geometries investigated, and especially the linear 12-electrode Paul trap, exhibit an extended region where the trapping field almost vanishes. The amplitude of the field rises abruptly when approaching the trap electrodes, which leads to stable trapping along with the occurrence of planar and volume structures in a layer of about a few millimeters thick, which practically spans the inner electrode space. We also report particle oscillations around equilibrium positions, where gravity is balanced by the trapping potential. Close to the trap center particles are almost {\em frozen}, which means they can be considered motionless due to the very low amplitude of their oscillation. The oscillation amplitude increases as one moves away from the trap center. We also report regions of dynamical stability for trapped charged microparticles located far away from the center of the trap, as shown in Fig. \ref{microplasmas}. The values of the specific charge ratios for the alumina microparticles range from $5.4 \times 10^{-4}$ C/kg to $0.13 \times 10^{-3}$ C/kg \cite{Stoican2008}. The measured microparticle density is around 3700 kg/m$^3$.    

The experimental results we report deal with trapping of microparticles in multipole linear ion (Paul) traps (LIT) operating in air, under Standard Ambient Temperature and Pressure (SATP) conditions. We suggest such traps can be used to levitate and study different microscopic particles, aerosols and other constituents or polluting agents which might exist in the atmosphere. The research performed is based on previous results and experience \cite{Gheorghe98, Stoican2008, Mihalcea2008, We2, Visan2013, Lapitsky2015b}. Moreover, the paper brings new evidence with respect to recent work of the authors \cite{Vasilyak2013}. Numerical simulations were run in order to characterize microparticle dynamics.

\section{Physical modelling and computer simulation}\label{model}

In addition to experimental work, numerical simulation of charged particle dynamics was carried out, under conditions close to the experiment. Brownian dynamics has been used in order to study charged microparticle motion and thus identify regions of stable trapping. Numerical simulations take into account stochastic forces of random collisions with neutral particles, viscosity of the gas medium, regular forces produced by the a.c. trapping voltage and the gravitational force. Thus, microparticle dynamics is characterized by a stochastic Langevin differential equation \cite{Vasilyak2013, We2}: 

\begin{equation}
\label{eq.1}
m_p \frac{d^2 r}{dt^2} = F_t(r)-6 \pi \eta r_p \frac{dr}{dt} + F_b + F_g
\end{equation}
where $m$ and $r_p$ represent the microparticle mass and radius vector, $\eta$ is the dynamic viscosity of the gas medium with $\eta = 18.2 \:\mu$Pa$\cdot$s, and $F_t(r)$ is the ponderomotive force. The $F_b$ term stands for the stochastic delta-correlated forces accounting for stochastic collisions with neutral particles, while $F_g$ is the gravitational force. We have considered a microparticle mass density value $\rho_p = 3700$ kg/cm$^3$ \ \cite{Stoican2008, Visan2013}. In order to solve the stochastic differential equation~(\ref{eq.1}), the numerical method developed in \cite{Skeel2002} was used. 

The average Coulomb force acting on a microparticle owing to the contribution of each trap electrode can be expressed as the vector sum of forces of point-like charges uniformly distributed along the electrodes, as demonstrated \cite{Vasilyak2013, Lapitsky2015b}:

\begin{equation}
\label{eq.2}
|F_t(r)|=\sum\limits_s\frac{L U q}{2 N \ln{\left(\frac{R_2}{R_1}\right)}(r_s - r)^2},
\end{equation}
where $L$ is the length of the trap electrodes, $U$ is the trapping voltage: $V_{ac} \sin(\Omega t)$ or $V_{ac} \sin(\Omega t+\pi)$, $q$ is the microparticle charge, $N$ is the number of point-like charges for each trap electrode, $R_2$ and $R_1$ represent the radii of the grounded cylindrical shell surrounding the trap and trap electrode respectively, while $r$ and $r_s$ denote the vectors for microparticle and point-like charge positions respectively. Numerical simulations were run considering the following trap parameters: length of electrodes $L = 6.5$ cm, $V_{ac} = 2$ kV, $R_2 = 25$ cm, $R_1 = 3$ mm, and a trap radius value $r_t = 2$ cm. For such model the results of the computations depend on $\Phi_p$,  defined as
$$\Phi_p = \frac{V_{ac} \ q}{2 \ln{\left(\frac{R2}{R1}\right)}} .$$

\begin{figure}[bth]
    \begin{minipage}[h]{0.45\linewidth}
    \center{\includegraphics[width=\linewidth]{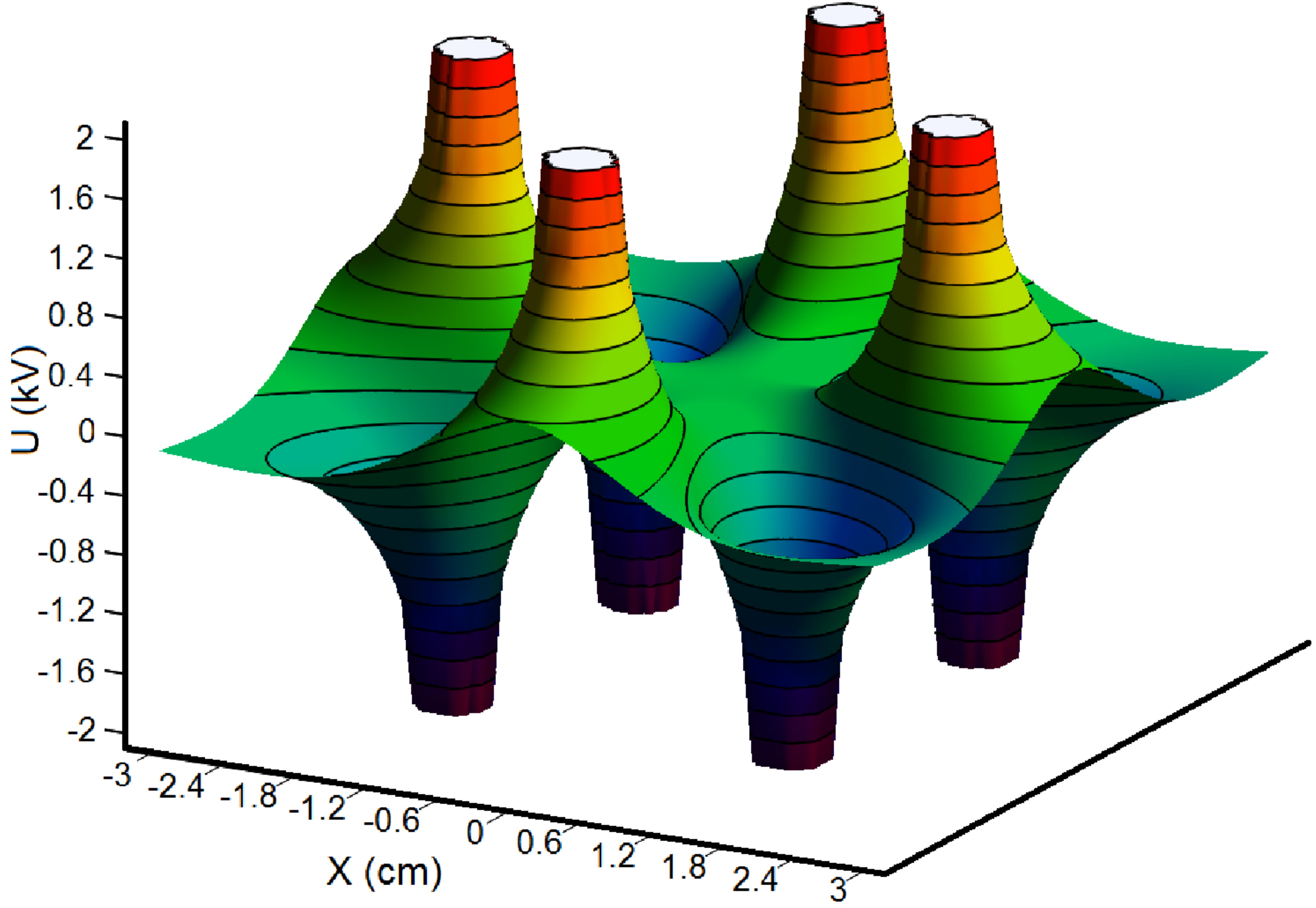}\\a)}
    \end{minipage}
    \begin{minipage}[h]{0.45\linewidth}
    \center{\includegraphics[width=\linewidth]{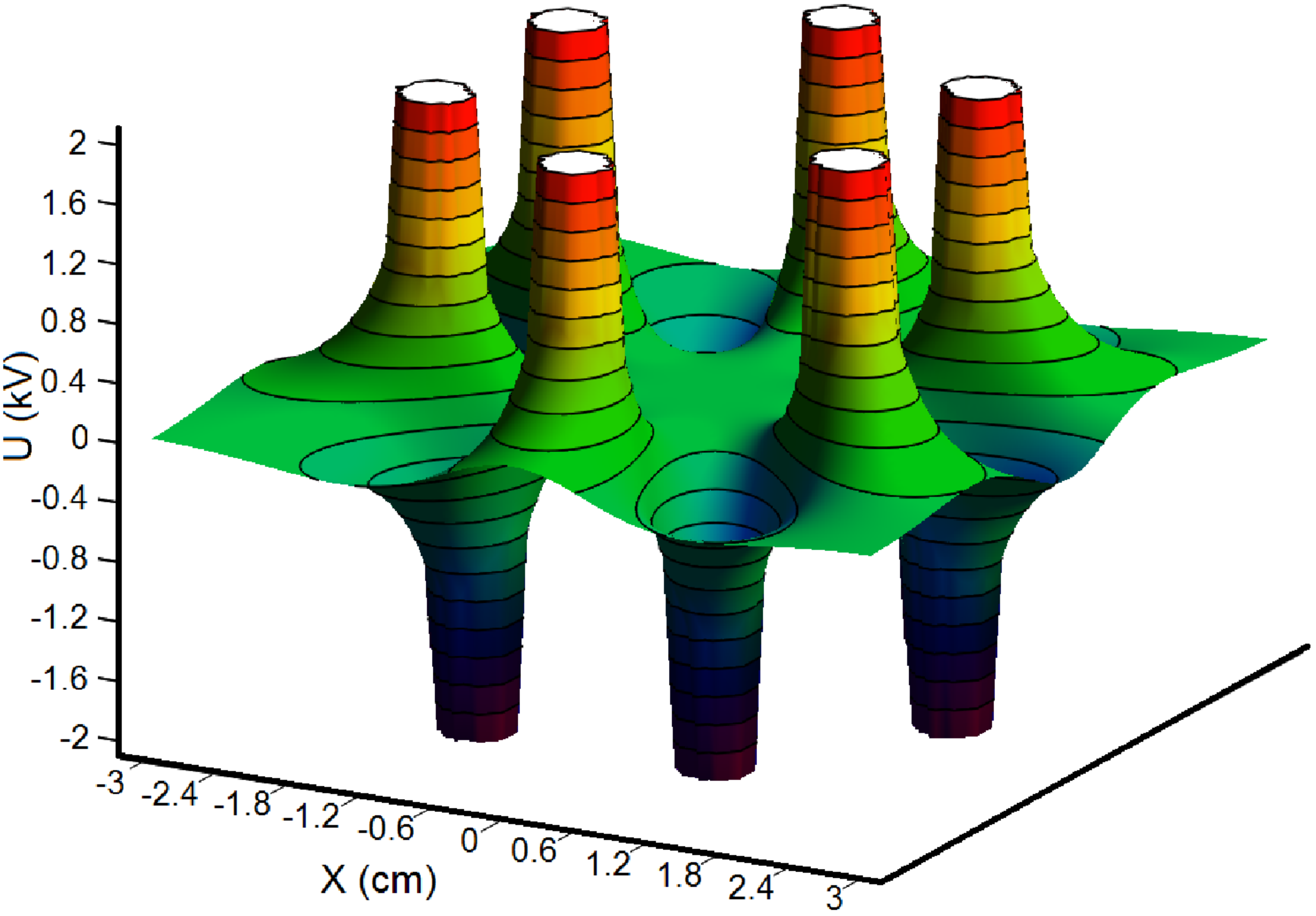}\\b)}
    \end{minipage}
    \caption{3D plots for the 8-electrode (a) and 12-electrode (b)  traps.} 
    \label{surf}
\end{figure}

The cross sections of the equipotential surfaces for the 8-electrode and 12-electrode linear trap are shown in Fig. \ref{surf}. The equation which characterizes the trap potential can be expressed as:
\begin{eqnarray}
 U \left(x, y\right) = \nonumber\\
  = \sum\limits_{j=1}^{N_{el}}\sum\limits_{s}\frac{\left(-1\right)^j L U/N}{\ln{\left(\frac{R_2}{R_1}\right)^2}\sqrt{\left(x - r_t\cos\left(\frac{2\pi j}{N_{el}}\right)\right)^2 + \left(y -r_t\sin\left(\frac{2\pi j}{N_{el}}\right)\right)^2+{z_s}^2}} \ .
\end{eqnarray}

We found that regions of stable microparticle confinement depend on the a.c. voltage, trap volume, number of trapped microparticles, average interparticle distance and consequently, on the repulsive forces produced by interparticle interactions defined by the particle charge $q$. To further minimize the influence of these physical factors, we used a higher value of the a.c. trapping voltage electrode $V_{ac} = 2$ kV. In such case, the electric charge of the captured particles will be lower and results of the simulations will be more or less universal.  

In Fig. \ref{surf} hills correspond to potential barriers and pits correspond to potential wells that attract microparticles. White holes inside the hills correspond to the cross section of the trap electrodes. Every half cycle of the a.c. voltage barriers and wells swap positions and each charged particle oscillates between them. Such a physical mechanism of particle oscillation results in dynamic, stable confinement \cite{Paul1990}. 

\begin{figure}[bth]
\begin{center}  
\includegraphics[width=0.5\linewidth]{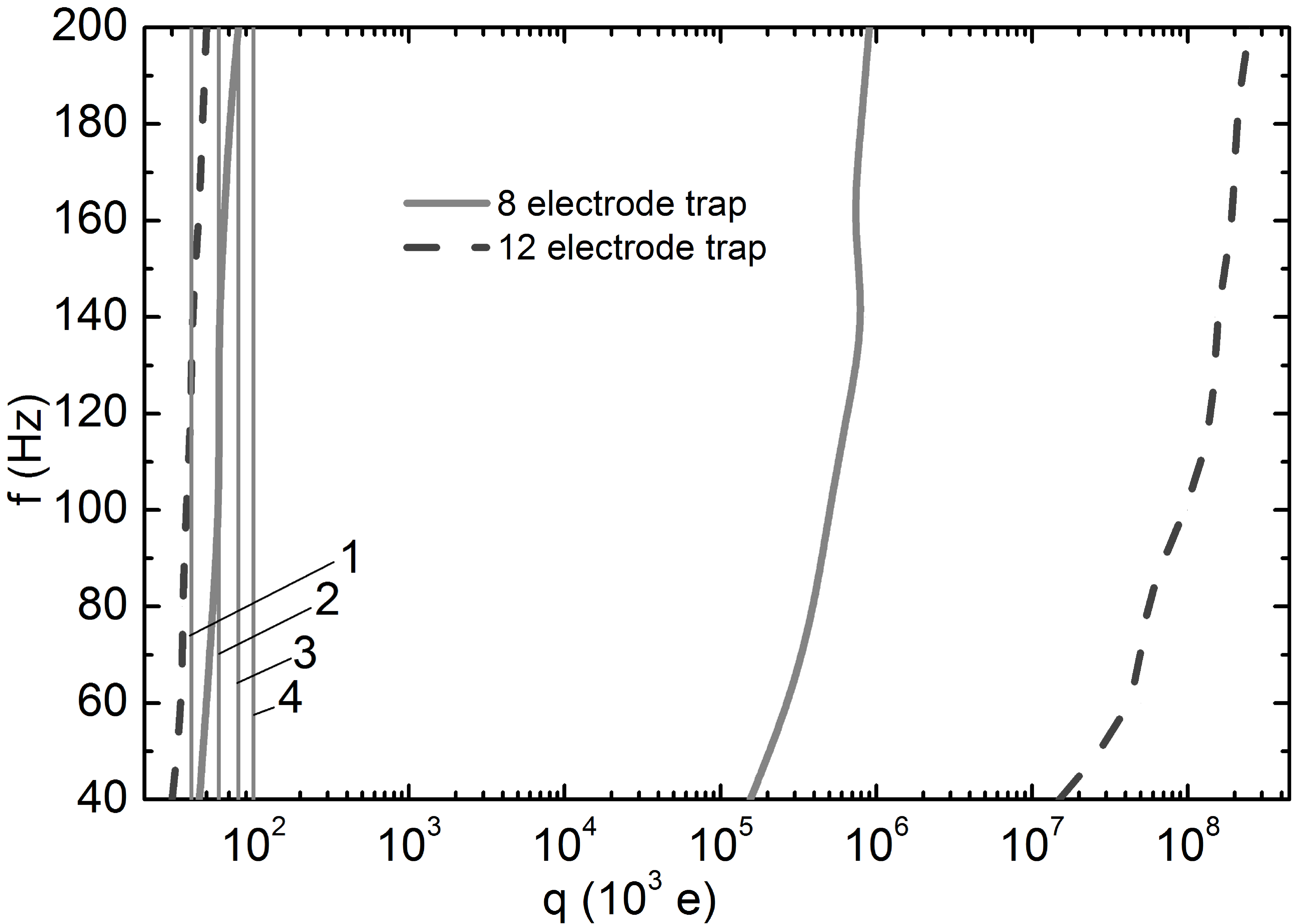}
\end{center}
\caption{The regions for single particle confinement in air, depending on the frequency $f$ of the a.c. voltage and electric charge $q$. Numerical simulations used the following characteristics: microparticle radius  $r_p = 5\: \mu$m and electrical charge ranging between $q = 3\cdot10^4 e$ to $5\cdot 10^{11} e$. Vertical lines 1 -- 4 correspond to electric charge values $q = 6, 8, 10, 12 \cdot10^4 e$ we have considered in order to estimate oscillation amplitudes within the trap.}
\label{8-12}
\end{figure}

Figure \ref{8-12} illustrates the trapping regions for a charged microparticle in an 8-electrode and 12-electrode trap, respectively. The confinement regions are located between correlated lines: for an 8-electrode trap between gray lines and for 12-electrode trap between black dash lines. Outside these regions traps cannot confine particles. For small values of the particle charge, the trap a.c. field cannot compensate the gravity force and particles flow through the trap. When the particle electric charge is large enough, the trap field is strong enough to push the microparticles out of the trap.

\begin{figure}[bth]
    \begin{minipage}[h]{0.38\linewidth}
    \center{\includegraphics[width=\linewidth]{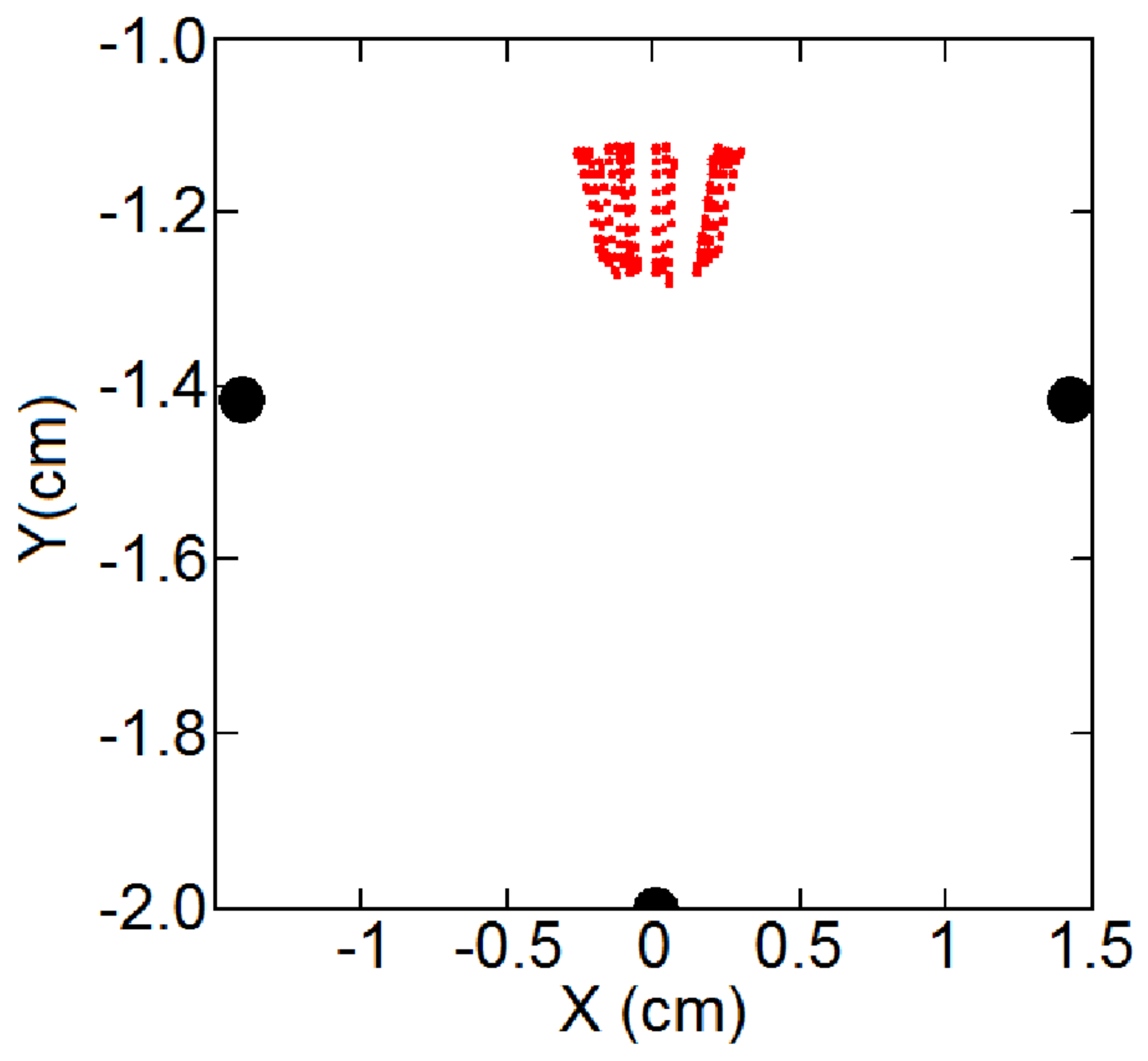}\\a)}
    \end{minipage}
    \begin{minipage}[h]{0.38\linewidth}
    \center{\includegraphics[width=\linewidth]{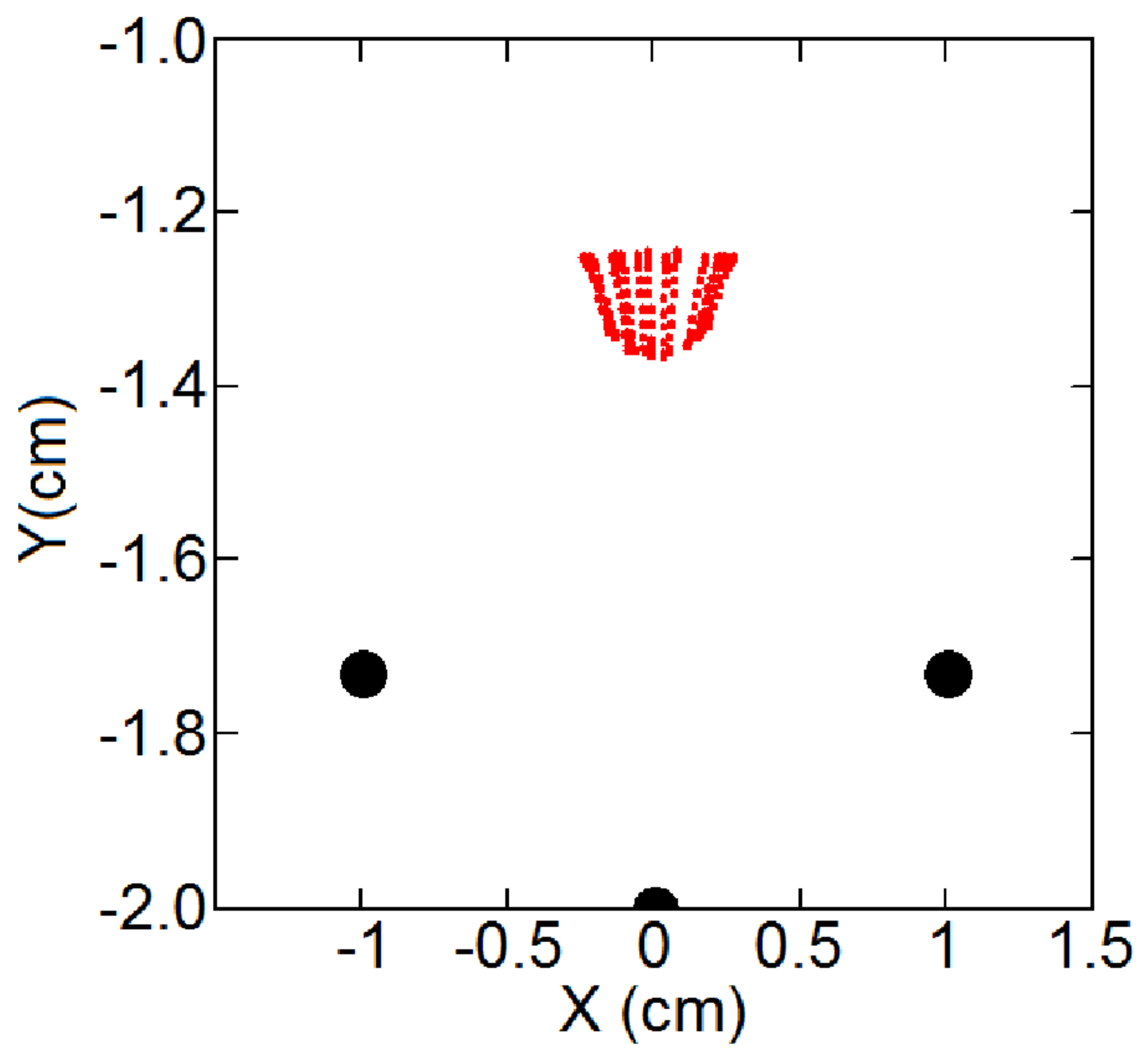}\\b)}
    \end{minipage}
    \caption{End views of the microparticle tracks in 8-electrode (a) and 12-electrode (b) traps at $f = 60$ Hz. Big black dots correspond to trap electrodes (not in scale). The microparticle electric charge value was chosen $q = 8 \cdot 10^4 e$. The d.c. voltage was not considered in the simulations.}
    \label{tracks8-12}
\end{figure}

To study the influence of the number of trap electrodes on the stability of alumina (dust) particles, we have investigated the average amplitude of particle oscillations. We considered the dynamics of a number of 20 microparticles confined within the trap while averaging the amplitudes of particle oscillations. To achieve that, we have chosen a period of time large enough (around ten periods of the a.c. trapping voltage) as to obtain stable particle oscillations. In order to evaluate the average oscillation amplitude, the amplitudes of all 20 particles were averaged. Trajectories for 20 particles confined in 8- and respectively 12-electrode traps, are shown in Fig. \ref{tracks8-12}. Particle trajectories are shifted downwards in the 12-electrode trap with respect to an 8-electrode trap, owing to a smaller gradient of the a.c. electric field.

\begin{figure}[bth]
    \begin{minipage}[h]{0.45\linewidth}
    \center{\includegraphics[width=\linewidth]{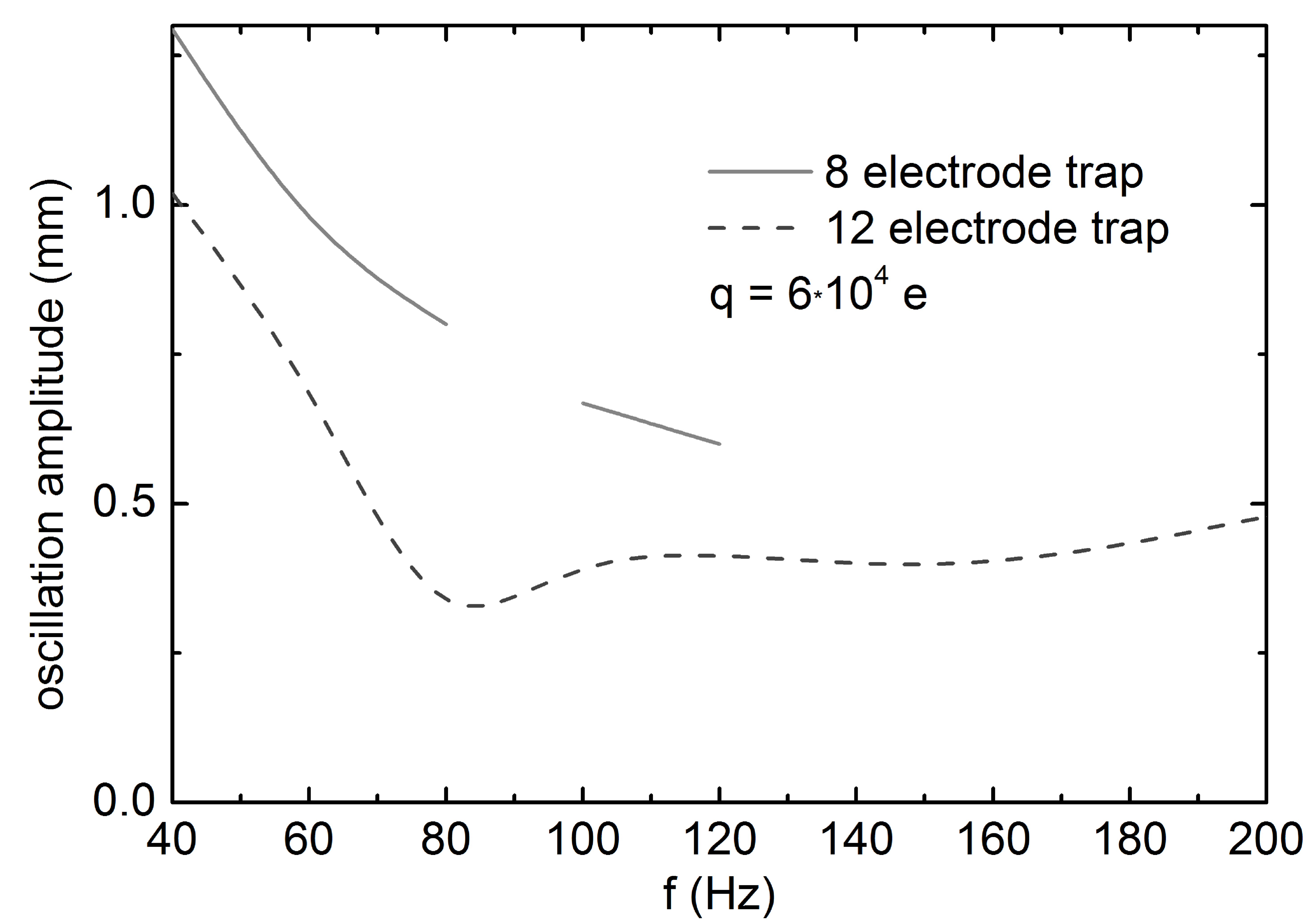}\\a)}
    \end{minipage}
    \begin{minipage}[h]{0.45\linewidth}
    \center{\includegraphics[width=\linewidth]{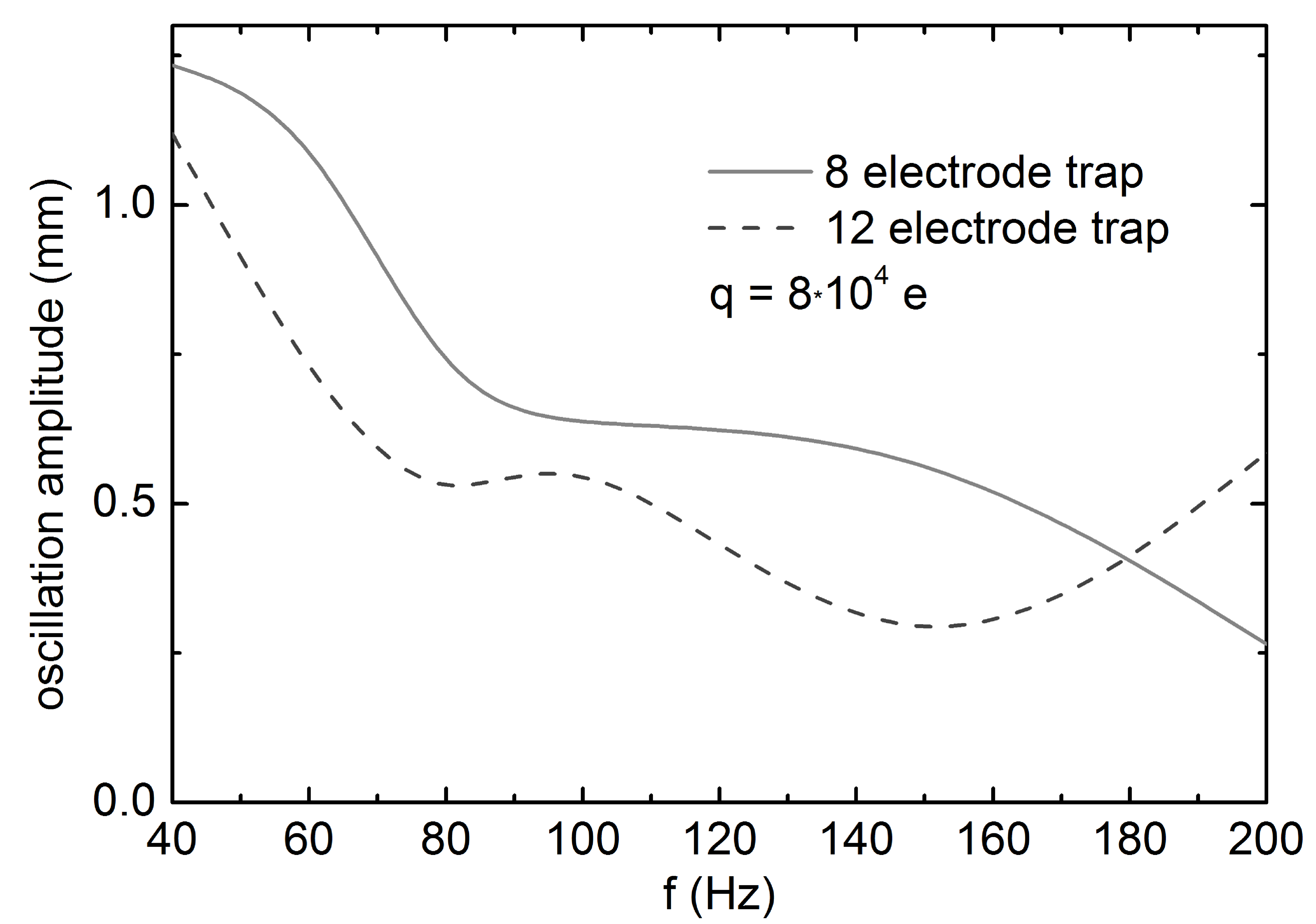}\\b)}
    \end{minipage}
    \begin{minipage}[h]{0.45\linewidth}
    \center{\includegraphics[width=\linewidth]{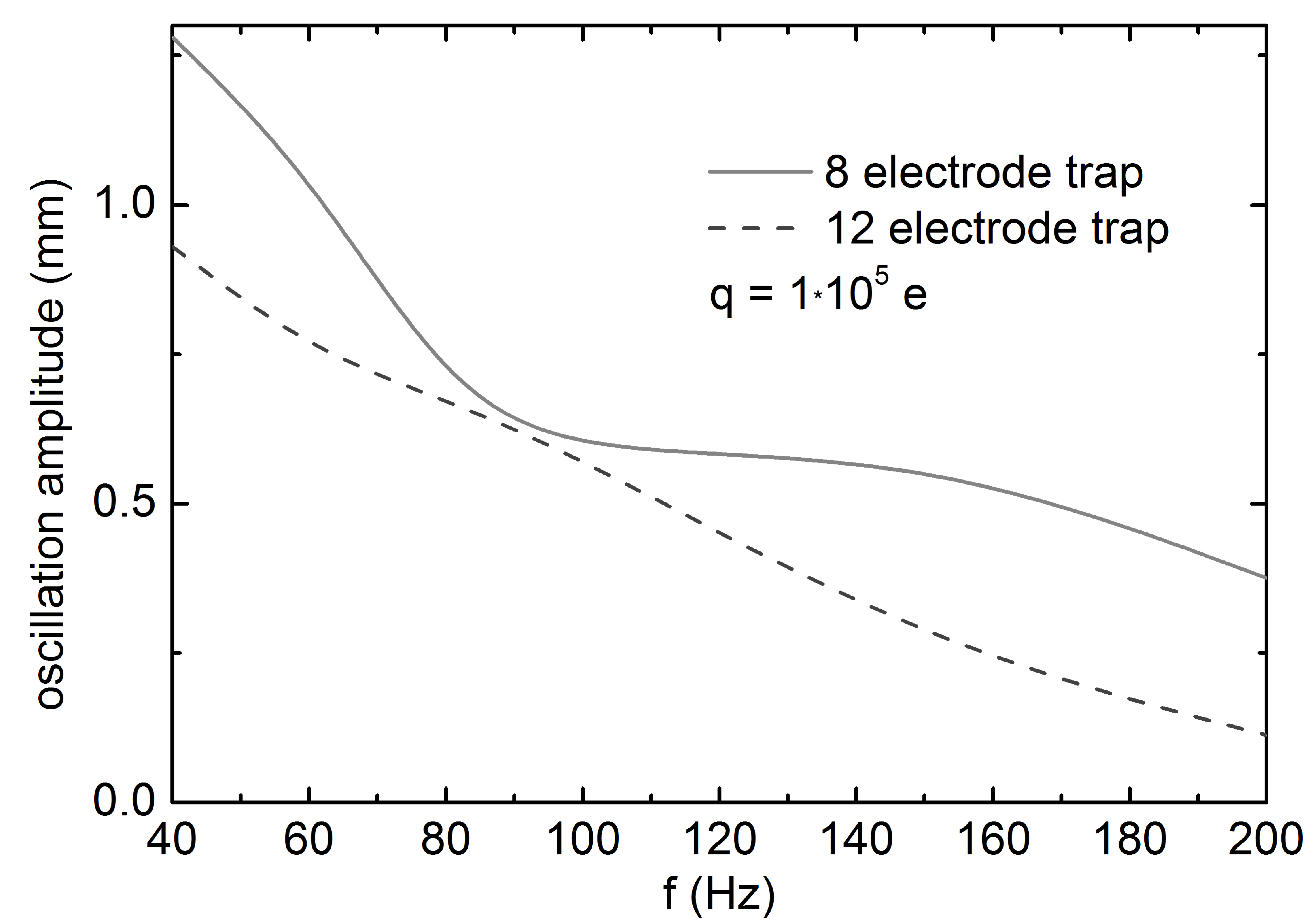}\\c)}
    \end{minipage}
    \begin{minipage}[h]{0.45\linewidth}
    \center{\includegraphics[width=\linewidth]{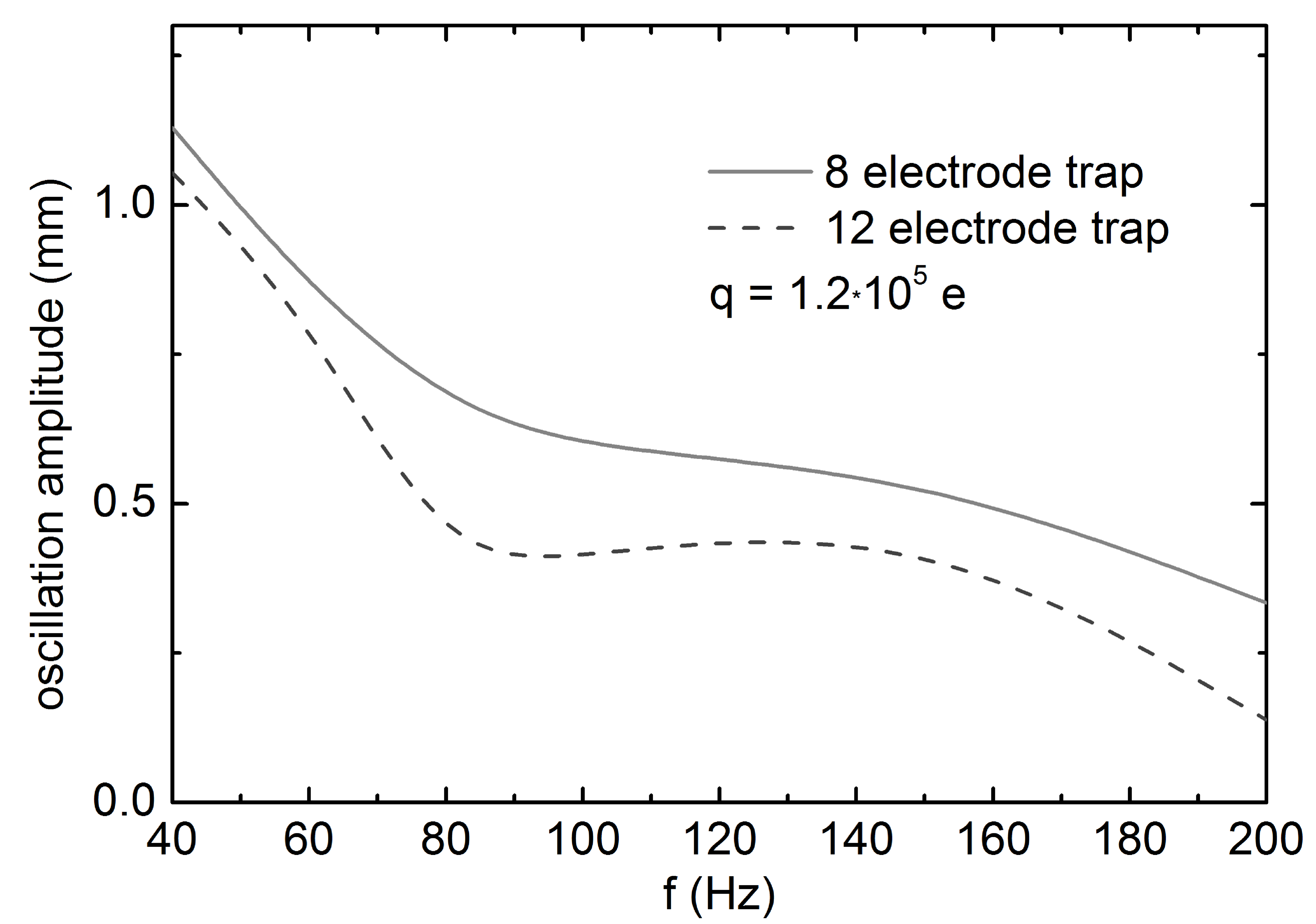}\\d)}
    \end{minipage}
    \caption{Dependency of average oscillation amplitude on the number of trap electrodes and particle charge $q$. Numerical simulations were performed for electric charge values ranging between $q = 6\cdot10^4 e$ to $1.2\cdot10^5 e$.}
    \label{ampl1}
\end{figure}

The dependence between particle oscillation amplitude and the number of trap electrodes is shown in Fig.~\ref{ampl1}. Such dependence is complex as it also depends on the particle charge. The physical reason explaining non-monotonic decay of some dependences on frequency and break in Fig.~\ref{ampl1} a) for an 8-electrode trap, might be associated with the occurrence of resonance effects. For example, in an 8-electrode trap a resonance effect is found at $\sim 80\div100$ Hz, for particles with an electric charge value of about $q = 8\cdot 10^4 e$. In such case the particles are pushed out of the trap, despite of the fact that all parameters are characteristic to the confinement region (Fig.~\ref{8-12}). As it follows from Fig.~\ref{8-12} for fixed particle charge values, the dependence of the averaged oscillation amplitude on the frequency in Fig.~\ref{ampl1} a) is beyond the confinement region when the frequency reaches a value of 120 Hz. Other pictures in Fig.~\ref{ampl1} b) - d) describe confinement regions with frequency values ranging from 40 up to 200 Hz.

\section{Conclusions}\label{conclusions}

The paper suggests use of Multipole Microparticle Electrodynamic Ion Traps (MMEITs) for trapping aerosols and microparticles, while bringing new experimental evidence on the advantages associated to such geometries. The mathematical model we propose and the numerical simulations performed simply create a clear and thorough picture of the phenomena we investigate. The pictures in \cite{Libbrecht2015} are similar to those we have obtained, apart from the fact that we have achieved such phenomena in multipole traps, under better stability and less sensibility to environment fluctuations (flows of air). The 12 electrode trap exhibits a variable geometry, which is one of the unique features that characterize the linear Paul traps we have tested since 1996. We emphasize on the fact that the stability region for multipole Paul traps is larger with respect to a quadrupole trap (electrodynamic balance configuration), a feature which we demonstrate by charting the electric field map within the trap using the electrolytic tank method, validated by experimental observations and numerical simulations.

The trap geometries investigated are characterized by more regions of stable trapping, some of them located close to the trap electrodes. Photos taken illustrate such phenomenon. By comparing the experimental data recorded, we can ascertain that stronger confinement, increased stability and larger microparticle numbers are achieved under conditions of dynamical stability in case of a 12-electrode trap. We also probe that the trap geometries investigated and especially the linear 12-electrode Paul trap, exhibit an extended region where the trapping field almost vanishes. The amplitude of the field rises abruptly in the vicinity of the trap electrodes, which leads to stable trapping and the appearance of planar and volume structures in a layer of about a few millimeters thick. Particle oscillations around equilibrium positions occur, where gravity is balanced by the trapping potential. Close to the trap center particles are almost {\em frozen}, which means that they can be considered as motionless due to the very low amplitude of their oscillation. The oscillation amplitude increases as one moves away from the trap center. Regions of dynamical stability are also observed for trapped charged microparticles located far away with respect to the trap center.

Experimental data recorded are backed up by numerical simulations results, which allows us to ascertain that a multipole trap exhibits a very extended region where the a.c. field intensity is weak, compared to a quadrupole trap. Due to the fact that the specific charge of the microparticles is very different compared to the specific charge of an ion, we suggest that miniaturized multipole trap geometries might be suited for ions trapped in ultrahigh vacuum conditions. Microparticles are thermalized by air friction or air drag, an effect similar with cooling of ions by collisions with neutral background molecules or sympathetic cooling. In spite of that, microparticle energy is still high and a lot of them are lost when loading the trap. For ions confined in multipole traps under ultrahigh vacuum conditions, there exists a large region where the trapping field is effectively zero. This drastically minimizes perturbing effects among which the foremost is the second-order Doppler shift, making these traps very suitable for high-resolution spectroscopy, quantum optics, quantum metrology and quantum information processing (QIP) experiments \cite{Knoop2015, Orszag2016, Burt2006}.

Brownian dynamics has been used in order to study charged microparticle motion and thus identify regions of stable trapping. A novel model is suggested for multipole Paul traps, as there were no other previous models to characterize such traps up to date. We have investigated microparticle dynamics in multipole traps by considering a Langevin differential equation, which takes into account stochastic forces of random collisions with neutral particles, viscosity of the gas medium, regular forces produced by the a.c. trapping voltage and the gravitational force. Regions of stable particle trapping have been identified by experimental observations, in good agreement with the numerical simulations results. To study the influence of the number of trap electrodes on the dynamics and stability of alumina (dust) particles we have confined, the average amplitude of particle oscillations in the trap was investigated. Numerical simulations clearly emphasize the influence of resonance effects on the confinement region.

MMEITs represent versatile tools that enable study complex Coulomb systems (microplasmas). The traps under test operate in air, at Standard Ambient Temperature and Pressure (SATP) conditions. The microplasmas we have studied provide a basis for the study of microparticle dynamics and phenomena associated to such kind of experiments, as well as the appearance of ordered structures, crystal like formations, corresponding to phase transitions. Such traps can be adapted to work in ultrahigh vacuum conditions. Moreover, an ion trap can be coupled to an Aerosol Mass Spectrometer to investigate atmospheric aerosol (nano)particles, which makes them very well suited for environment monitoring studies, coupled with other techniques such as LIDAR.

The most prominant advantage of multipole traps  lies in the fact that the spatial extent of the region of confinement expands with the number of electrodes, and our results strongly support this hypothesis. And it is an aspect of utmost importance, as recent highly accurate absolute frequency measurements of several microwave and optical frequency standards have yielded sensitive probes of possible temporal and spatial changes of some of the fundamental constants. In these and other experiments, greater sensitivity would be possible using atomic frequency standards with inaccuracy below 1 part in $10^{15}$ typical of the best, present-day atomic standards. This motivates the quest for new ion trap geometries, with an increased signal-to-noise ratio (SNR).

\section{Acknowledgements}
\indent
The authors would like to acknowledge support provided by the Ministery of Education, Research and Inovation from Romania (ANCS-National Agency for Scientific Research), contracts PN09.39.03.01, UEFISCDI Contract No. 90/06.10.2011, and ROSA Contract No. 53/19. 11. 2013.

Mathematical model and Brownian dynamics simulations have been carried out in the Joint Institute for High Temperatures, Moscow, Russian Academy of Sciences (RAS), under financial support by the Russian Foundation for Basic Research (grant No. 15-08-02835).

\section{References}

\bibliographystyle{unsrt}
\bibliography{Multipole}

\begin{thebibliography}{10}

\bibitem{Davidson2001}
Ronald~C. Davidson.
\newblock {\em Physics of Nonneutral Plasmas}.
\newblock Imperial College Press \& World Scientific, 2001.

\bibitem{Werth2005a}
G{\"u}nther Werth.
\newblock Non-neutral plasmas and collective phenomena in ion traps.
\newblock In Andreas Dinklage, Thomas Klinger, Gerrit Marx, and Lutz
  Schweikhard, editors, {\em Plasma Physics: Confinement, Transport and
  Collective Effects, Lecture Notes in Physics}, volume 670, 2005.

\bibitem{Tsyto2008}
Vadim~N. Tsytovich, Gregory~E. Morfill, Sergey~V. Vladimirov, and Hubertus~M.
  Thomas.
\newblock {\em Elementary Physics of Complex Plasmas}, volume 731 of {\em
  Lecture Notes in Physics}.
\newblock Springer, Berlin Heidelberg, 1st edition, 2008.

\bibitem{Mendonca2013}
J.~T. Mendon\c{c}a and Hugo Ter\c{c}as.
\newblock {\em Physics of Ultra-Cold Matter: Atomic Clouds, Bose-Einstein
  Condensates and Rydberg Plasmas}.
\newblock Springer, 2013.

\bibitem{Ott2014}
T.~Ott, M.~Bonitz, L.~G. Stanton, and M.~S. Murillo.
\newblock Coupling strength in coulomb and yukawa one-component plasmas.
\newblock {\em Phys. Plasmas}, 21:113704, 2014.

\bibitem{Knoop2015}
Martina Knoop, Irene Marzoli, and Giovanna Morigi, editors.
\newblock {\em Ion Traps for Tomorrow Applications}, volume Course 189 of {\em
  Proc. Int. School of Physics Enrico Fermi}.
\newblock IOS Press and Societ\`a Italiana di Fisica, Bologna, Italy, 2015.

\bibitem{Bonitz2010a}
Michael Bonitz, Patrick Ludwig, and Norman Horing, editors.
\newblock {\em Introduction to Complex Plasmas}, volume~59 of {\em Springer
  Series on Atomic, Optical and Plasma Physics}.
\newblock Springer-Verlag, Berlin Heidelberg, 1st edition, 2010.

\bibitem{Fortov2010}
Vladimir~E. Fortov and Gregor~E. Morfill, editors.
\newblock {\em Complex and Dusty Plasmas: From Laboratory to Space}.
\newblock CRC Press, 2010.

\bibitem{Shukla2002}
Padma~Kant Shukla and A.~A. Mamun.
\newblock {\em Introduction to Dusty Plasma Physics}.
\newblock IOP Publishing Ltd, 2002.

\bibitem{Morfill2009}
Gregor~E. Morfill and Alexei~V. Ivlev.
\newblock Complex plasmas: An interdisciplinary research field.
\newblock {\em Rev. Mod. Phys.}, 81:1353--1404, 2009.

\bibitem{Bonitz2014}
Michael Bonitz, Jose Lopez, Kurt Becker, and Hauke Thomsen, editors.
\newblock {\em Complex Plasmas: Scientific challenges and Technological
  Opportunities}, volume~82 of {\em Springer Series on Atomic, Optical and
  Plasma Physics}.
\newblock Springer, Springer International Publishing Switzerland, 1st edition,
  2014.

\bibitem{Shukla2002b}
Padma~Kant Shukla.
\newblock {\em Dust Plasma Interaction in Space}.
\newblock Nova Science Publishers, 2002.

\bibitem{Trevi2006}
Adam~J. Trevitt.
\newblock {\em Ion Trap Studies of Single Microparticles: Optical Resonances
  and Mass Spectroscopy}.
\newblock PhD thesis, School of Chemistry, The University of Melbourne, 2006.

\bibitem{Bonitz2010b}
Michael~M. Bonitz, C.~Henning, and D.~Block.
\newblock Complex plasmas: a laboratory for strong correlations.
\newblock {\em Rep. Prog. Phys.}, 73:066501, 2010.

\bibitem{Davis2002}
E.~James Davis and Gustav Schweiger.
\newblock {\em The Airborne Microparticle: Its Physics, Chemistry, Optics, and
  Transport Phenomena}.
\newblock Springer-Verlag, Berlin Heidelberg, 1st edition, 2002.

\bibitem{Libbrecht2015}
Kenneth~G. Libbrecht and Eric~D. Black.
\newblock Improved microparticle electrodynamic ion traps for physics teaching.
\newblock \url{http://newtonianlabs.com/itdemo/iontraps.pdf/}, 2015.

\bibitem{Fortov2007}
Vladimir~E. Fortov, Igor~T. Iakubov, and Alexey~G. Khrapak.
\newblock {\em Physics of Strongly Coupled Plasma}.
\newblock International Series of Monographs on Physics. Clarendon Press,
  Oxford, 2007.

\bibitem{Dubin1999}
Daniel H.~E. Dubin and T.~M. O'Neil.
\newblock Trapped nonneutral plasmas, liquids, and crystals (the thermal
  equilibrium states).
\newblock {\em Rev. Mod. Phys.}, 71:87--172, 1999.

\bibitem{Lisin2013}
Evgeny~A. Lisin, Rinat~A. Timirkhanov, Olga~S. Vaulina, Oleg~F. Petrov, and
  Vladimir~E. Fortov.
\newblock Influence of external perturbations on the interaction between grains
  in plasma.
\newblock {\em New J. Phys.}, 15:053004, 2013.

\bibitem{Chen2007}
Goong Chen, David~A. Church, Berthold-Georg Englert, Carsten Henkel, Bernd
  Rohwedder, Marlan~O. Scully, and M.~Suhail Zubairy.
\newblock {\em Quantum Computing Devices: Principles, Design, and Analysis}.
\newblock CRC Press, 2007.

\bibitem{Bolli2003}
J.~J. Bollinger, J.~M. Kriesel, T.~B. Mitchell, L.~B. King, M.~J. Jensen, W.~M.
  Itano, and D.~H.~E. Dubin.
\newblock Laser-cooled ion plasmas in penning traps.
\newblock {\em J. Phys. B}, 36:499--510, 2003.

\bibitem{Major2005}
Fouad~G. Major, Viorica~N. Gheorghe, and G{\"u}nther Werth.
\newblock {\em Charged Particle Traps: Physics and Techniques of Charged
  Particle Field Confinement}, volume~37 of {\em Springer Series on Atomic,
  Optical and Plasma Physics}.
\newblock Springer-Verlag, Berlin Heidelberg, 1st edition, 2005.

\bibitem{Haroche2013}
Serge Haroche and Jean-Michel Raimond.
\newblock {\em Exploring the Quantum: Atoms, Cavities and Photons}.
\newblock Oxford University Press, 2013.

\bibitem{Quint2014}
Wolfgang Quint and Manuel Vogel, editors.
\newblock {\em Fundamental Physics in Particle Traps}, volume 256 of {\em
  Springer Tracts in Modern Physics}.
\newblock Springer, Springer-Verlag, Berlin Heidelberg, 2014.

\bibitem{Knoop2014}
Martina Knoop, Niels Madsen, and Richard~C. Thompson, editors.
\newblock {\em Physics with Trapped Charged Particles: Lectures from the Les
  Houches Winter School}.
\newblock Imperial College Press, 2014.

\bibitem{Paul1958}
Wolfgang Paul, H.~P. Reinhard, and U.~von Zahn.
\newblock Das elektrische massenfilter als massenspektrometer und
  isotopentrenner.
\newblock {\em Z. f. Physik}, 152:143--182, 1958.

\bibitem{Paul1990}
Wolfgang Paul.
\newblock Electromagnetic traps for charged and neutral particles.
\newblock {\em Rev. Mod. Phys.}, 62:531--540, 1990.

\bibitem{Ghosh1995}
Pradip~K. Ghosh.
\newblock {\em Ion Traps}.
\newblock Clarendon Press, 1995.

\bibitem{Werth2009}
G{\"u}nther Werth, Viorica~N. Gheorghe, and Fouad~G. Major.
\newblock {\em Charged Particle Traps II: Applications}, volume~54 of {\em
  Springer Series on Atomic, Optical and Plasma Physics}.
\newblock Springer-Verlag, Berlin Heidelberg, 1st edition, 2009.

\bibitem{Wuerker1959}
Ralph~F. Wuerker, Haywood Shelton, and Robert~V. Langmuir.
\newblock Electrodynamic containment of charged particles.
\newblock {\em J. Appl. Phys}, 30:342--349, 1959.

\bibitem{Winter1991}
H.~Winter and H.~W. Ortjohann.
\newblock Simple demonstration of storing macroscopic particles in a paul trap.
\newblock {\em J. Appl. Phys}, 59:807--813, 1991.

\bibitem{Izmailov1995}
Alexander~F. Izmailov, Stephen Arnold, Stephen Holler, and Allan~S. Myerson.
\newblock Microparticle driven by parametric and random forces: Theory and
  experiment.
\newblock {\em Phys. Rev. E}, 52:1325--1332, 1995.

\bibitem{Mihalcea2010}
Bogdan~M. Mihalcea and Gina G.~Vi\c san.
\newblock Nonlinear ion trap stability analysis.
\newblock {\em Phys. Scr.}, page 014057, 2010.

\bibitem{Akerman2010}
N.~Akerman, S.~Kotler, Y.~Glickman, Y.~Dallal, A.~Keselman, and R.~Ozeri.
\newblock A single-ion nonlinear mechanical oscillator.
\newblock {\em Phys. Rev. A}, 82:061402(R), 2010.

\bibitem{Blumel1990}
R.~Bl{\"u}mel, E.~Peik, W.~Quint, and H.~Walther.
\newblock Phase transitions of stored laser-cooled ions.
\newblock {\em Acta Phys. Polonica A}, 78:419--432, 1990.

\bibitem{Walther1995}
H.~Walther.
\newblock From a single ion to a mesoscopic system - crystallization of ions in
  paul traps.
\newblock {\em Phys. Scr.}, T59:360--368, 1995.

\bibitem{Schlipf1996}
Stefan Schlipf.
\newblock From a single ion to a mesoscopic system: Crystalization of ions in
  paul traps.
\newblock In Alain Aspect, W.~Barletta, and R.~Bonifacio, editors, {\em
  Coherent and Collective Interactions of Particles and Radiation Beams},
  volume 131 of {\em Proceedings of the International School of Physics Enrico
  Fermi}, 1996.

\bibitem{Margolis2004}
H.~S. Margolis, G.~P. Barwood, G.~Huang, H.~A. Klein, S.~N. Lea, K.~Szymaniec,
  and P.~Gill.
\newblock Hertz-level measurement of the optical clock transition in a single
  $^{88}$sr$^+$ ion.
\newblock {\em Science}, 306:1355--1358, 2004.

\bibitem{Rosen2007}
T.~Rosenband, P.~O. Schmidt, D.~B. Hume, W.~M. Itano, T.~M. Fortier, J.~E.
  Stalnaker, K.~Kim, S.~A. Diddams, J.~C.~J. Koelemeij, J.~C. Bergquist, and
  D.~J. Wineland.
\newblock Observation of the $^1$s$_0 \rightarrow ^3$p$_0$ clock transition in
  $^27$al$_+$.
\newblock {\em Phys. Rev. Lett.}, 98:220801, 2007.

\bibitem{Bushev2013}
P.~Bushev, G.~H\'{e}tet, L.~Slodi\v{c}ka, D.~Rotter, M.~A. Wilson,
  F.~Schmidt-Kaler, J.~Eschner, and R.~Blatt.
\newblock Shot-noise-limited monitoring and phase-locking of the motion of a
  single trapped ion.
\newblock {\em Phys. Rev. Lett}, 110:133602, 2013.

\bibitem{Leibf2003}
D.~Leibfried, R.~Blatt, C.~Monroe, and D.~Wineland.
\newblock Quantum dynamics of single trapped ions.
\newblock {\em Rev. Mod. Phys.}, 75:281--324, 2003.

\bibitem{Home2011}
J.~P. Home, D.~Hanneke, J.~D. Jost, D.~Leibfried, and D.~J. Wineland.
\newblock Normal modes of trapped ions in the presence of anharmonic trap
  potentials.
\newblock {\em New J. Phys.}, 13:073026, 2011.

\bibitem{Peik2006}
Ekkehard Peik, Tobias Schneider, and Christian Tamm.
\newblock Laser frequency stabilization to a single ion.
\newblock {\em J. Phys. B: At. Mol. Opt. Phys.}, 39:145--158, 2006.

\bibitem{March2010}
Raymond~E. March and John F.~J. Todd.
\newblock {\em Practical Aspects of Ion Trap Mass Spectrometry}, volume~5.
\newblock CRC Press, Boca Raton, 1st edition, 2010.

\bibitem{Ludlow2015}
Andrew~D. Ludlow, Martin~M. Boyd, Jun Ye, Ekkehard Peik, and Piet O. Schmidt.
\newblock Optical atomic clocks.
\newblock {\em Rev. Mod. Phys.}, 87:637--701, 2014.

\bibitem{Chiav2005}
J.~Chiaverini, R.~B. Blakestad, J.~Britton, J.~D. Jost, C.~Langer,
  D.~Leibfried, R.~Ozeri, and D.~J. Wineland.
\newblock Surface-electrode architecture for ion-trap quantum information
  processing.
\newblock {\em Quantum Information and Computation}, 5:419--439, 2005.

\bibitem{Leibf2007}
Diedrich Leibfried, David~J. Wineland, R.~Brad Blakestad, John~J. Bollinger,
  Joseph Britton, John Chiaverini, R.~J. Epstein, Wayne~M. Itano, John~D. Jost,
  Emanuel Knill, Christopher Langer, Roee Ozeri, Rainer Reichle, Signe
  Seidelin, N.~Shiga, and J.~H. Wesenberg.
\newblock Towards scaling up trapped ion quantum information processing.
\newblock {\em Hyperfine Interactions}, 174:1--7, 2007.

\bibitem{Zutic2004}
Igor \v{Z}uti\'{c}, Jaroslav Fabian, and S.~Das Sarma.
\newblock Spintronics: Fundamentals and applications.
\newblock {\em Rev. Mod. Phys.}, 76:323--410, 2004.

\bibitem{Kim2005}
Jungsang Kim, S.~Pau, Z.~Ma, H.~R. McLellan, J.~V. Gates, A.~Kornblit, R.~M.
  Jopson, I.~Kang, M.~Dinu, and Richart~E. Slusher.
\newblock System design for large-scale ion trap quantum information processor.
\newblock {\em Quantum Inf. Comput.}, 5:515--537, 2005.

\bibitem{King2012}
Julian~A. King, John~K. Webb, Michael~T. Murphy, Victor~V. Flambaum, Robert~F.
  Carswell, Matthew~B. Bainbridge, Michael~R. Wilczynska, and F.~Elliot Koch.
\newblock Spatial variation in the fine-structure constant new results from
  {VLT/UVES}.
\newblock {\em Mon. Not. R. Astron. Soc.}, 422:3370--3414, 2012.

\bibitem{Pandis1995}
Spyros~N. Pandis, Anthony~S. Wexler, and John~S. Seinfeld.
\newblock Dynamics of tropospheric aerosols.
\newblock {\em J. Phys. Chem.}, 99:9646--9659, 1995.

\bibitem{Seinfeld2006}
John~H. Seinfeld and Spyros~N. Pandis.
\newblock {\em Atmospheric Chemistry and Physics: From Air Pollution to Climate
  Change}.
\newblock Wiley, 2006.

\bibitem{Draine2003}
Bruce~T. Draine.
\newblock Interstellar dust grains.
\newblock {\em Annual Review of Astronomy and Astrophysics}, 41:241--289, 2003.

\bibitem{Fortov2011}
Vladimir~E. Fortov.
\newblock {\em Extreme States of Matter}.
\newblock Springer-Verlag, 2011.

\bibitem{Kulkarni2011}
Pramod Kulkarni, Paul~A. Baron, and Klaus Willeke, editors.
\newblock {\em Aerosol Measurement: Principles, Techniques and Applications}.
\newblock Wiley, 2011.

\bibitem{Wester2009}
Roland Wester.
\newblock Radiofrequency multipole traps: tools for spectroscopy and dynamics
  of cold molecular ions.
\newblock {\em J. Phys. B: At. Mol. Opt. Phys.}, 42:154001, 2009.

\bibitem{Kaiser1991}
Raymond E.~Kaiser Jr., R.~Graham Cooks, George C.~Stafford Jr., John E.~P.
  Syka, and Philip~H. Hemberger.
\newblock Operation of a quadrupole ion trap mass spectrometer to achieve high
  mass/charge ratios.
\newblock {\em Int. J. Mass Spectr.}, 106:79--115, 1991.

\bibitem{Schwartz1991}
Jae~C. Schwartz, John E.~P. Syka, and Ian Jardine.
\newblock High resolution on a quadrupole ion trap mass spectrometer.
\newblock {\em J. Am. Soc. Mass Spectrometry}, 2:198--204, 1991.

\bibitem{March2005}
Raymond~E. March and John F.~J. Todd.
\newblock {\em Quadrupole Ion Trap Mass Spectrometry}, volume 165 of {\em
  Chemical Analysis: A Series of Monographs on Analytical Chemistry and its
  Applications}.
\newblock Wiley, Hoboken, New Jersey, 2nd edition, 2005.

\bibitem{Nie2008}
Zongxiu Nie, Fenping Cui, Minglee Chu, Chung-Hsuan Chen, Huan-Cheng Chang, and
  Yong Cai.
\newblock Calibration of a frequency-scan quadrupole ion trap mass spectrometer
  for microparticle mass analysis.
\newblock {\em Int. J. Mass Spectr.}, 270:8--15, 2008.

\bibitem{Trevi07}
Adam~J. Trevitt, Philip~J. Wearne, and Evan~J. Bieske.
\newblock Calibration of a quadrupole ion trap for particle mass spectrometry.
\newblock {\em Int. J. Mass Spectrometry}, 262:241--246, 2007.

\bibitem{Smith2008}
Trevor~A. Smith, Adam~J. Trevitt, Philip~J. Wearne, Evan~J. Bieske, Lachlan~J.
  McKimmie, and Damian~K. Bird.
\newblock Morphology-dependent resonance emission from individual micron sized
  particles.
\newblock In M.~N. Berberan-Santos, editor, {\em Fluorescence of
  Supermolecules, Polymers, and Nanosystems}, volume~4 of {\em Springer Series
  on Fluorescence}, pages 415--429, Berlin Heidelberg, 2008. Springer.

\bibitem{Trevi09}
Adam~J. Trevitt, Philip~J. Wearne, and Evan~J. Bieske.
\newblock Coalescence of levitated polystyrene microspheres.
\newblock {\em J. Aerosol Science}, 40:431--438, 2009.

\bibitem{Stoican2008}
Ovidiu~S. Stoican, Lauren{\c t}iu~C. Dinc{\u a}, Gina Vi{\c s}an, and {\c
  S}tefan R{\u a}dan.
\newblock Acoustic detection of the parametrical resonance effect for a
  one-component microplasma consisting of the charged microparticles stored in
  the electrodynamic traps.
\newblock {\em J. Opt. Adv. Mat.}, 10:1988--2000, 2008.

\bibitem{Wang2006}
Shenyi Wang, Christopher~A. Zordan, and Murray~V. Johnston.
\newblock Chemical characterization of individual, airborne sub-10 nm particles
  and molecules.
\newblock {\em Anal. Chem.}, 78:1750--1754, 2006.

\bibitem{Kurten2007}
Andreas K{\"u}rten, Joachim Curtius, Frank Helleis, Edward~R. Lovejoy, and
  Stephan Borrmann.
\newblock Development and characterization of an ion trap mass spectrometer for
  the on-line chemical analysis of atmospheric aerosol particles.
\newblock {\em Int. J. Mass Spectr.}, 265:30--39, 2007.

\bibitem{Gerlich2008a}
Dieter Gerlich.
\newblock The study of cold collisions using ion guides and traps.
\newblock In {\em Low Temperatures and Cold Molecules}.
\newblock pages 121--174, Imperial College Press, 2008.

\bibitem{Gerlich2008b}
Dieter Gerlich.
\newblock The production and study of ultra-cold molecular ions.
\newblock In {\em Low Temperatures and Cold Molecules}.
\newblock pages 295--344, Imperial College Press, 2008.

\bibitem{Gerlich1992}
Dieter Gerlich.
\newblock Inhomogeneous electrical radio-frequency fields: A versatile tool for
  the study of processes with slow ions.
\newblock {\em Adv. Chem. Phys.}, 82:1--176, 1992.

\bibitem{Trippel2006}
S.~Trippel, J.~Mikosch, R.~Berhane, Rico Otto, Matthias Weidem{\"u}ller, and
  Roland Wester.
\newblock Photodetachment of cold ${OH}^{-}$ in a multipole ion trap.
\newblock {\em Phys. Rev. Lett.}, 97:193003, 2006.

\bibitem{Otto2009}
R.~Otto, P.~Hlavenka, S.~Trippel, J.~Mikosch, K.~Singer, M.~Weidem{\"u}ller,
  and Roland Wester.
\newblock How can a 22-pole ion trap exhibit ten local minima in the effective
  potential?
\newblock {\em J. Phys. B: At. Mol. Opt. Phys.}, 42:154007, 2009.

\bibitem{Burt2006}
E.~A. Burt and R.~L. Tjoelker.
\newblock Sub-$10^{-16}$ frequency stability in multi-pole linear ion trap
  standards: Reduction of number-dependent sensitivity.
\newblock In Joseph~H. Yuen, editor, {\em IPN Progress Report}, volume 42-166,
  pages 1--15, Pasadena, 08 2006. NASA, Jet Propulsion Lab.

\bibitem{Vasilyak2013}
Leonid~M. Vasilyak, Vladimir~I. Vladimirov, Lidiya~V. Deputatova, Dmitriy~S.
  Lapitsky, Vladimir~I. Molotkov, Vladimir~Yakovlevich Pecherkin, Vladimir~S.
  Filinov, and Vladimir~E. Fortov.
\newblock Coulomb stable structures of charged dust particles in a dynamical
  trap at atmospheric pressure in air.
\newblock {\em New Journal of Physics}, 15:043047, 2013.

\bibitem{Orszag2016}
Miguel Orszag.
\newblock {\em Quantum Optics: Including Noise Reduction, Trapped Ions, Quantum
  Trajectories, and Decoherence}.
\newblock Springer Intl. Publishing, 2016.

\bibitem{Vinitsky2015}
Eugene~A. Vinitsky, Eric~D. Black, and Kenneth~G. Libbrecht.
\newblock Particle dynamics in damped nonlinear quadrupole ion traps.
\newblock {\em Am. J. Phys.}, 83:313--319, 2015.

\bibitem{Riehle2004}
Fritz Riehle.
\newblock {\em Frequency Standards: Basics and Applications}, volume 256.
\newblock Wiley, Wiley-VCH Verlag, Weinheim, 2004.

\bibitem{Champenois2009}
Caroline Champenois.
\newblock About the dynamics and thermodynamics of trapped ions.
\newblock {\em J. Phys. B: At. Mol. Opt. Phys.}, 42:154002, 2009.

\bibitem{Beranek2010}
M.~Ber{\'a}nek, I.~{\v C}erm{\'a}k, Z.~N{\v e}me{\v c}ek, and J.~{\v
  S}afr{\'a}nkov{\'a}.
\newblock Trapping charged microparticles in the linear quadrupole trap.
\newblock In {\em WDS'10 Proceedings of Contributed Papers: Part II – Physics
  of Plasmas and Ionized Media}.

\bibitem{Lapitsky2016}
D.~S. Lapitsky.
\newblock Micro-particle charge determination using a linear paul trap with the
  end electrode.
\newblock {\em J. Phys.: Conf. Series}, 666:012026.1--4, 2016.

\bibitem{Deputatova2015a}
L.~V. Deputatova, V.~S. Filinov, D.~S. Lapitsky, V.~Ya. Pecherkin, R.~A.
  Syrovatka, L.~M. Vasilyak, and V.~I. Vladimirov.
\newblock Measurement of the charge of a single dust particle.
\newblock {\em J. Phys.: Conf. Series}, 653:012129.1--5, 2015.

\bibitem{Lapitsky2015b}
D.~S. Lapitsky.
\newblock Effective forces and pseudopotential wells and barriers in the linear
  paul trap.
\newblock {\em J. Phys: Conference Series}, 653:012130, 2015.

\bibitem{Lapitsky2015a}
D.~S. Lapitsky, V.~S. Filinov, L.~M. Vasilyak, R.~A. Syrovatka, L.~V.
  Deputatova, V.~I. Vladimirov, and V.~Ya. Pecherkin.
\newblock Confinement of the charged microparticles by alternating electric
  fields in a gas flow.
\newblock {\em Eur. Phys. Lett.}, 110, 2015.

\bibitem{Visan2013}
Gina~Vi\c san and Ovidiu Stoican.
\newblock An experimental setup for the study of the particles stored in an
  electrodynamic linear trap.
\newblock {\em Rom. J. Phys.}, 58:171--180, 2013.

\bibitem{Prestage1999}
J.~D. Prestage, R.~L. Tjoelker, and L.~Maleki.
\newblock Higher pole linear traps for atomic clock applications.
\newblock In {\em Proceedings of the 1999 Joint Meeting of the European
  Frequency and Time Forum and the IEEE International Frequency Control
  Symposium}, volume~1, 1999.

\bibitem{Jusko2012}
P.~Jusko, {\v S}.~Rou{\v c}ka, R.~Pla{\v s}il, D.~Gerlich, and J.~Glos{\'i}k.
\newblock Electron spectrometer multipole trap: First experimental results.
\newblock In {\em WDS'12 Proceedings of Contributed Papers: Part II – Physics
  of Plasmas and Ionized Media}.

\bibitem{Roucka2009}
{\v S}.~Rou{\v c}ka, A.~Podolnik, P.~Jusko, P.~Kotr{\'i}k, R.~Pla{\v s}il, and
  J.~Glos{\'i}k.
\newblock Combination of a 22-pole trap with an electron energy filter - study
  of associative detachment ${H}^{-} + {H} \rightarrow {H}_{2} + {e}^{-}$.
\newblock In {\em WDS'09 Proceedings of Contributed Papers: Part II – Physics
  of Plasmas and Ionized Media}.

\bibitem{Roucka2010}
{\v S}.~Rou{\v c}ka, A.~Podolnik, P.~Jusko, P.~Kotr{\'i}k, R.~Pla{\v s}il, and
  J.~Glos{\'i}k.
\newblock Study of capture and cooling of ${H}^{-}$ ions in rf octopole with
  superimposed magnetic field.
\newblock In {\em WDS'10 Proceedings of Contributed Papers: Part II – Physics
  of Plasmas and Ionized Media}.

\bibitem{Clark2013}
Robert~J. Clark.
\newblock Ideal multipole ion traps from planar ring electrodes.
\newblock {\em Appl. Phys. B: Lasers and Optics}, volume113, pages 171--178, 2013.

\bibitem{Cartarius2013}
Florian Cartarius, Cecilia Cormick, and Giovanna Morigi.
\newblock Stability and dynamics of ion rings in linear multipole traps.
\newblock {\em Phys. Rev. A}, 87:013425, 2013.

\bibitem{Champenois2013}
C.~Champenois, J.~Pedregosa-Gutierrez, M.~Marciante, D.~Guyomarc'h, and
  M.~Houssin.
\newblock A double ion trap for large coulomb crystals.
\newblock In {\em AIP Conf. Proc.}, volume 1521, pages 210--219, 2013.

\bibitem{Gheorghe98}
Viorica~N. Gheorghe, Liviu Giurgiu, Ovidiu Stoican, Drago{\c s} Cacicovschi,
  Radu Molnar, and Bogdan Mihalcea.
\newblock Ordered structures in a variable length {AC} trap.
\newblock {\em Acta Physica Polonica A}, 93:625--629, 1998.

\bibitem{Mihalcea2008}
Bogdan~M. Mihalcea, Gina T.~Vi\c san, Liviu~C. Giurgiu, and {\c S}tefan R{\u
  a}dan.
\newblock Optimization of ion trap geometries and of the signal to noise ratio
  for high resolution spectroscopy.
\newblock {\em J. Opt. Adv. Mat.}, 10:1994--1998, 2008.

\bibitem{We2}
V.~S. Filinov, D.~S. Lapitsky, L.~V. Deputatova, L.~M. Vasilyak, V.~I.
  Vladimirov, and O.~A. Sinkevich.
\newblock {\em Contrib. Plasma Phys.}, 52:66--69, 2012.

\bibitem{Skeel2002}
R.~D. Skeel and J.~A. Izaguirre.
\newblock {\em Molecular Physics}, 100:3885--3891, 2002.

\end{thebibliography}

\end{document}